\documentclass[10pt,journal,doublecolumn,twoside]{IEEEtran}

\ifCLASSOPTIONcompsoc
\usepackage[nocompress]{cite}
\else
\usepackage{cite}
\fi

\ifCLASSINFOpdf

\else

\fi
\usepackage{amsmath,lipsum}
\usepackage{cuted}
\usepackage{graphicx}
\usepackage{epstopdf}
\usepackage{amsmath}
\usepackage{subeqnarray}
\usepackage{cases}
\usepackage{algorithmic}
\usepackage{algorithm}
\usepackage{array}
\usepackage{multicol}
\usepackage{stfloats}

\usepackage{url}
\usepackage{harpoon}
\usepackage{booktabs}
\usepackage{multirow}
\usepackage{amsfonts}
\usepackage{xcolor}
\usepackage{tabularx}
\usepackage{amssymb,varwidth}
\usepackage{bm}
\usepackage{float}
\usepackage{graphicx}
\usepackage{subfigure}
\newenvironment{proof}{{\indent \it Proof:\quad}}{\hfill $\blacksquare$\par}

\newtheorem{proposition}{Proposition}

\begin{document}

\title{Environment-Aware AUV Trajectory Design and Resource Management for Multi-Tier Underwater Computing}
\author{Xiangwang Hou, \emph{Student Member, IEEE,}
        Jingjing Wang, \emph{Senior Member, IEEE},  Tong Bai, \emph{Member, IEEE}, \\ Yansha Deng, \emph{Member, IEEE},  Yong Ren, \emph{Senior Member, IEEE}, Lajos Hanzo, \emph{Life Fellow, IEEE}
\thanks{This work of Jingjing Wang was supported in part by the National Natural Science Foundation of China under grant No. 62071268 and grant No. 6222101, in part by the Young Elite Scientist Sponsorship Program by the China Association for Science and Technology under Grant No. 2020QNRC001, and in part by the Fundamental Research Funds for the Central Universities. T. Bai was supported in part by the National Natural Science Foundation of China under Grant 62101015.  Y. Deng was partially supported by Engineering and Physical Sciences Research Council (EPSRC), U.K., under Grant EP/W004348/1. Y. Ren was supported in part by the National Natural Science Foundation of China under grant No. 62127801, in part by the National Key R \&D Program of China under Grant 2020YFD0901000, and in part by the project `The Verification Platform of Multi-tier Coverage Communication Network for Oceans (LZC0020)' of Peng Cheng Laboratory. Moreover, L. Hanzo would like to acknowledge the financial support of the Engineering and Physical Sciences Research Council projects EP/W016605/1 and EP/P003990/1 (COALESCE) as well as of the European Research Council's Advanced Fellow Grant QuantCom (Grant No. 789028). \textit{(Corresponding author: Jingjing Wang.)}}
\thanks{X. Hou is with the Department of Electronic Engineering, Tsinghua University, Beijing, 100084, China. (E-mail: xiangwanghou@163.com.)}
\thanks{J. Wang and T. Bai are with the School of Cyber Science and Technology, Beihang University, Beijing 100191, China. (E-mail: drwangjj@buaa.edu.cn, tongbai@buaa.edu.cn.)}
\thanks{Y. Deng is with the Department of Engineering, King's College London, London WC2R 2LS, U.K. (E-mail: yansha.deng@kcl.ac.uk.)}
\thanks{Y. Ren  is with the Department of Electronic Engineering, Tsinghua University, Beijing, 100084, China, and also with the Network and Communication Research Center, Peng Cheng Laboratory, Shenzhen, 518055, China (E-mail: reny@tsinghua.edu.cn.)}
\thanks{L. Hanzo is with the School of Electronics and Computer Science, University of Southampton, Southampton, SO17 1BJ, UK. (E-mail: lh@ecs.soton.ac.uk.)}
}

\maketitle

\begin{abstract}
The Internet of underwater things (IoUT) is envisioned to be an essential part of maritime activities. Given the IoUT devices' wide-area distribution and constrained transmit power, autonomous underwater vehicles (AUVs) have been widely adopted for collecting and forwarding the data sensed by IoUT devices to the surface-stations. In order to accommodate the diverse requirements of IoUT applications, it is imperative to conceive a  multi-tier underwater computing (MTUC) framework by carefully harnessing both the computing and the communications as well as the storage resources of both the surface-station and of the AUVs as well as of the IoUT devices. Furthermore, to meet the stringent energy constraints of the IoUT devices and to reduce the operating cost of the MTUC framework, a joint environment-aware AUV trajectory design and resource management problem is formulated, which is a high-dimensional NP-hard problem. To tackle this challenge, we first transform the problem into a Markov decision process (MDP) and solve it with the aid of the asynchronous advantage actor-critic (A3C) algorithm. Our simulation results demonstrate the superiority of our scheme.
\end{abstract}
\section{Introduction}\label{AA}

As an extension of the Internet of things (IoT) in underwater environments, the Internet of underwater things (IoUT) is envisioned to be a crucial enabler for supporting diverse maritime activities \cite{9328873}. More explicitly, the IoUT aims for constructing a ``smart ocean" by connecting various underwater devices, e.g. sensors, robots, cameras, to monitor and reconstruct underwater objects and environments \cite{8863935}. In contrast to the terrestrial IoT systems, radio frequency (RF)-based techniques are unsuitable for the IoUT, owing to the severe absorption of electromagnetic waves in underwater environments. As a remedy, underwater acoustic communications (UAC) \cite{9399315,7102676} are widely adopted, but it still remains unrealistic for energy-limited IoUT devices to directly transmit their collected data to a surface-station through long-distance propagation, because ten-times higher transmit power is required compared to RF-based communications. To cope with this issue, autonomous underwater vehicles (AUV) have been widely adopted for data collection in underwater environments \cite{9356608,9647007}.


The seminal AUV-aided data collection techniques have routinely been based on a fixed AUV trajectory, such as an ellipse \cite{yoon2012aurp}. In this case, the IoUT devices distant from the AUV's trajectory have to aggregate their data at the IoUT devices in the close proximity of the AUV's trajectory for delivering it to AUVs. This inevitably leads to redundant communications and to potentially excessive energy requirements, especially at the data aggregation nodes. Hence, to overcome this impediment, recent studies opted for optimizing the AUV trajectory for actively collecting data from the IoUT devices \cite{han2019auv,duan2020value,gjanci2017path}. However, only the specific locations of the IoUT devices are considered in these research contributions, while ignoring the impact of hostile environmental factors, such as dynamically fluctuating water velocity, vortex, etc., which may lead to excessive propulsion energy consumption and even disable the AUV.

Apart from the data collector node mentioned above, AUVs may also play the role of an intermediate node for data relaying. However, the requirement of ocean exploration activities is not limited to communications. Besides sensors, a large number of advanced devices have been harnessed, such as diverse underwater robots. Consequently, a large variety of computing and storage tasks has to be processed in a time-sensitive manner. For example, when considering robots, their tasks have to be completed in time for adjusting the next mission. Although these devices are indeed equipped both with computing and storage capabilities, it is challenging to handle all the tasks locally, given their limited battery lives. Hence, it is beneficial to establish a multi-tier computing \cite{9676649} framework by integrating both the computing and the communications as well as storage resources of surface-stations and of AUVs, as well as of the devices for providing on-demand computing services.

Both AUV-centric \cite{8445615,9451536,9068243,8263193} and IoUT-centric \cite{9312959, gjanci2017path, 9548959} designs were considered in the open literature conceived either for latency-minimization or for energy-minimization. However, both types of designs have their limitations. As a remedy, we propose a system-level framework for maximizing the benefits of an intrinsically amalgamated hierarchical network comprised of IoUT devices, AUVs, and surface-stations. Note that it is not a simple conglomerate of its constituent components. For example, a rechargeable AUV and an IoUT device anchored underwater may consume the same energy but they have entirely different effects on the whole system, which deserves specific investigation.

Against this background, we design a multi-tier underwater computing (MTUC) framework intrinsically amalgamating both the computing and communications as well as storage resources of surface-stations and AUVs as well as IoUT devices for providing on-demand services for IoUT applications. Our new contributions are summarized as follows:


\begin{itemize}
  \item To the best of our knowledge, this is the first attempt to integrate the surface-stations, AUVs, and IoUT devices to form an MTUC framework for providing on-demand underwater computing services instead of simply collecting the sensory data for satisfying the diverse requirements of advanced IoUT applications.

   \item  Considering the limitations of both the AUV-centric and IoUT-centric designs, we conceive a system-level optimization model for maximizing the profits gleaned from the perspective of economics by integrating our environment-aware trajectory design, communication resource allocation, computation offloading and data caching.

  \item Since the problem formulated is NP-hard and high-dimensional, conventional methods cannot deal with it well. Hence, we transform it into a Markov decision process (MDP) and employ an asynchronous advantage actor-critic (A3C) algorithm \cite{9063667} for solving it.

  \item Our simulation results show that the proposed scheme is capable of improving the system's profit by relying on environment-aware trajectory design and always exhibits better convergence speed and scalability in the face of an escalating problem dimension than other state-of-the-art schemes.
\end{itemize}

The remainder of the paper is organized as follows. Section II reviews the related state-of-the-art. In Section III,  we describe the system model and formulate an optimization problem for maximizing the system's profit. Section IV introduces how we transform the optimization problem to an MDP and utilize A3C to solve it. In Section V, a range of experiments is carried out to show the efficiency of the proposed scheme. Section VI concludes the paper.

\section{Related work}
AUV-aided data collection has been extensively studied in recent years. Early efforts were focused on the collaborative transmissions of IoUT devices, while the trajectory of the AUV was usually assumed to be fixed \cite{yoon2012aurp,chen2012mobicast,khan2019auv}. As the first attempt to introduce AUV to relay the data of IoUT devices, Yoon \emph{et al.} \cite{yoon2012aurp} proposed a new underwater routing scheme, where the sensing devices send their data to an aggregation device either directly or via a multi-hop transmission, and then the aggregation device transmits the data aggregated to the AUV when it passes by. For reducing the number of communication hops, the sensing devices intelligently select the next hop according to the aggregation device's preference. With the objective of minimizing the energy consumption of the IoUT devices, Chen \emph{et al.} \cite{chen2012mobicast} conceived a novel routing protocol relying on the selective awake-sleep mechanism of IoUT devices and accurate estimation of the AUV's coverage range. Khan \emph{et al.} \cite{khan2019auv} investigated an energy-efficient AUV-assisted clustering scheme, comprised of a fixed time-slot-based intra-cluster communication mechanism with a wake-up sleep cycle and a sectoring mechanism, which is capable of reducing the processing latency, while avoiding excessive energy consumption. However, the previous contributions relying on data aggregation among IoUT devices impose an excessive burden on the aggregation devices selected.

For enhancing the battery lives, AUVs may be intelligently configured to cruise and collect the data of all the IoUT devices. The trajectory design of AUVs will be revealed to have a significant impact on the performance of the IoUT system, including both IoUT devices and the AUV. Hence, a series of treatises were dedicated to the AUV's trajectory design, where some of them aim for reducing the operating cost of the AUV in terms of cruising distance or energy consumption \cite{hollinger2012underwater,ma2012tour,7888568}. Others focused on whether the IoUT devices' requirements are satisfied~\cite{gjanci2017path, 9548959,9312959}. Specifically, considering the unreliable communications between the IoUT devices and AUVs,  Hollinger \emph{et al.} \cite{hollinger2012underwater} formulated a communication-limited data collection problem as a special traveling salesperson problem (TSP) and presented both an AUV path planning method as well as a communication protocol to solve it. The efficiency of the proposed strategy was validated both under a deterministic access and a random access scenario. To reduce the cruising distance of the AUV, Ma \emph{et al.} \cite{ma2012tour} designed a spanning tree covering algorithm for solving the path planning problem formulated. Faigl \emph{et al.} \cite{7888568} proposed employing a self-organizing map and an unsupervised learning technique to find a short path for the AUV considering the priority of the IoUT devices, which have a low computational complexity. This regime may also be readily extended to multi-AUV scenarios. With the emergence of advanced IoUT applications, such as mission-critical IoUTs \cite{9520368}, extremely stringent IoUT device requirements have to be considered. Hence, meeting these requirements of the IoUT applications with limited resources has drawn significant research attention. Bearing in mind that the value of the sensed data rapidly decays in time,  the authors proposed a heuristic adaptive greedy AUV path-finding algorithm to find an optimal path having the maximal data value delivered to the aggregation devices \cite{gjanci2017path}. Liu \emph{et al.} \cite{9548959} presented a hybrid data collection scheme, taking both the timeliness and energy efficiency requirements of IoUT devices into consideration. To guarantee the freshness of the collected data,  Fang \emph{et al.} \cite{9312959} introduced the concept of age of information and designed a two-stage algorithm for the joint optimization of the resource allocation and trajectory planning of AUV-aided IoUTs. At the time of writing, however, there is no recommendation in the open literature  for optimizing the amalgamated system's performance relying on integrating both the surface-station, AUVs, as well as the IoUT devices. It is beneficial to construct a system-level optimization framework for balancing the operating cost and meeting the IoUT devices' requirements. However there are some pioneering works on system-level optimization in terrestrial networks. Wang \emph{et al. }\cite{chenmeng2017} conceived a revenue-maximizing framework for cellular networks by jointly considering the computation offloading, resource allocation and content caching. Focusing on accuracy-aware machine learning (ML) tasks in the Internet of Industrial Things,  Fan \emph{et al. }\cite{9794317} constructed a long-term average system cost optimization framework by jointly considering the resources of sensors, edge server and cloud server, as well as the inference accuracy of the ML tasks. However, when the application scenario changes from cellular networks to AUV-aided underwater networks, the research mentioned above is no longer applicable. Hence the system considered deserves further study. Therefore, in this paper, a max-profit problem, integrating environment-aware trajectory design, communication resource allocation, computation offloading and data caching, is conceived for filling this knowledge gap.

\section{SYSTEM MODEL AND PROBLEM FORMULATION}\label{BB}
\subsection{Network Model}\label{A}
\begin{figure}
\centering
\includegraphics[width=8.7cm]{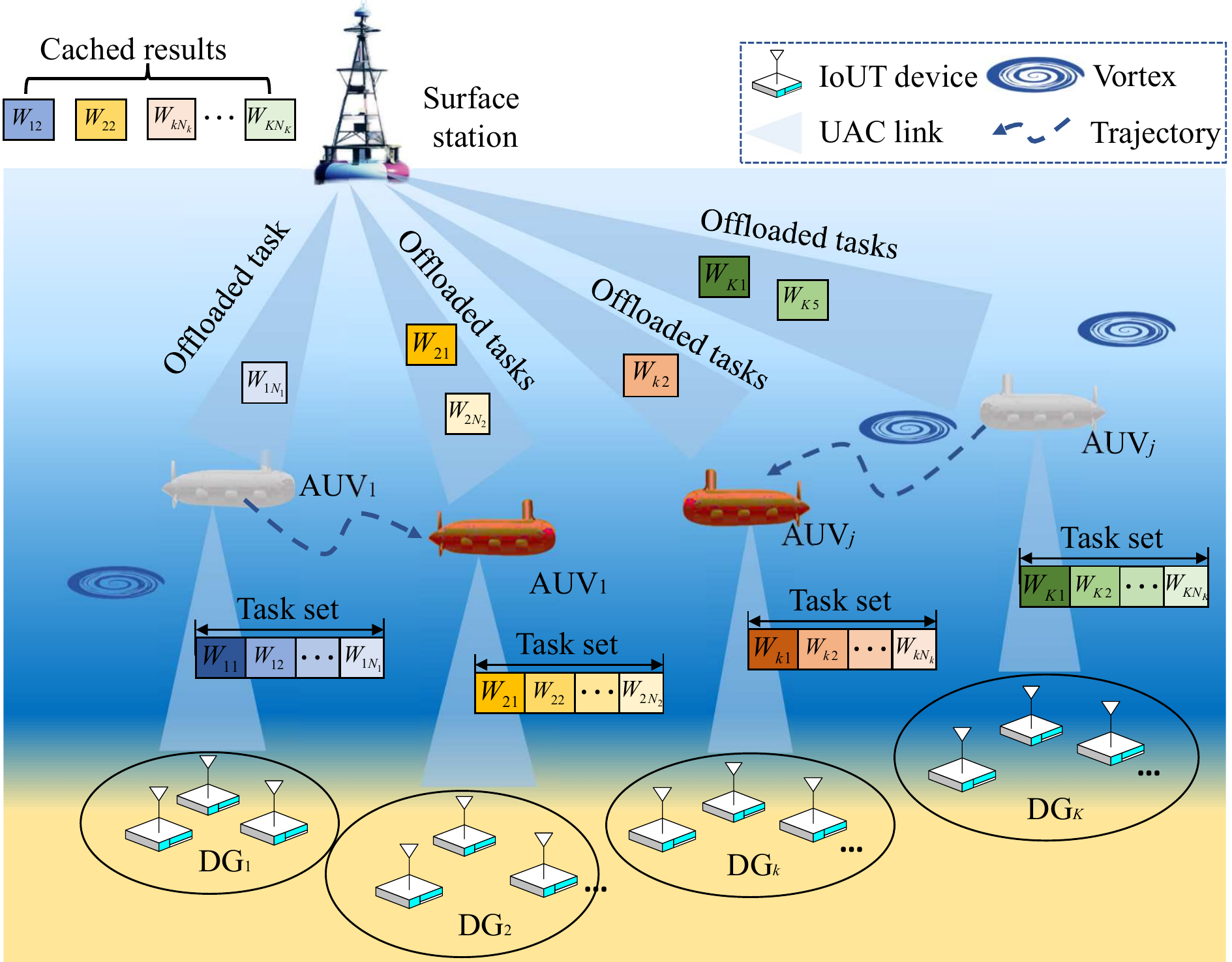}
\caption{The architecture of MTUC.}\label{SystemModel}
\vspace{-0.5cm}
\end{figure}
 Fig. \ref{SystemModel} shows our MTUC architecture, where multiple AUVs communicating with surface-stations perpetually cruising to provide computing service for a set of IoUT devices distributed in several device groups (DGs).  Each AUV starts from the point of origin sight below the surface-station, and supports the assigned DGs in turn. We assume that there is a single surface-station, $M$ AUVs, and $K$ DGs. The $M$ AUVs are denoted by the set $\boldsymbol{AUV}=\{AUV_1,  {AUV}_2,\ldots, {AUV}_M\}$, while the $K$ DGs are represented by the set $\boldsymbol{DG}=\{{{DG}}_1, {{DG}}_2,\ldots, {{DG}}_K\}$. Let us assume that there are a total of ${{N}_{k}}$ IoUT devices located in ${{DG}}_k$, which are represented by a set $\boldsymbol{{ND}_{k}}=\{{{n}_{{{k}_{1}}}},{{n}_{{{k}_{2}}}},\ldots,{{n}_{{{k}_{{N}_{k}}}}}\}$. For brevity, let $\boldsymbol{M}=\{1,  2,\ldots, M\}$ represent the subscript of the AUVs, while $\boldsymbol{K}=\{1,2,\ldots,K\}$ the subscript of the DGs, and $\boldsymbol{{N}_{k}}=\{1,2,\ldots,{N}_{k}\}$ as the subscript of the IoUT devices located in {DG}$_k$. Let  furthermore $\boldsymbol{P}_{\textrm{SS}}=(0,0,H)$, $\boldsymbol{P}_{j}^{\textrm{A}}=\left(x_{j}^{\textrm{A}}, y_{j}^{\textrm{A}}, d_{0}\right)$,  $\boldsymbol{P}_{k}^{\textrm{DG}}=\left(x_{k}^{\textrm{DG}}, y_{k}^{\textrm{DG}}, z_{k}^{\textrm{DG}} \right)$ and $\boldsymbol{P}_{ki}^{\textrm{S}}=\left(x_{ki}^{\textrm{S}}, y_{ki}^{\textrm{S}},h_{0} \right)$ represent the three-dimensional (3D) Euclidean coordinates of the surface-station, ${AUV}_k$,  ${{DG}}_k$, and  IoUT device ${{n}_{{{k}_{i}}}}$ located in ${{DG}}_k$, respectively.

We assume that each IoUT device has a task that has to be solved. The task generated by IoUT device ${{n}_{{{k}_{i}}}}$ can be represented by the twin tuple ${{W}_{{{k}_{i}}} \triangleq \{{Z}_{{{k}_{i}}},{\alpha}_{{{k}_{i}}}}\}$, where
${{Z}_{{{k}_{i}}}}$ represents the size of the input data (in bit), while ${{\alpha}_{{{k}_{i}}}}$ is the computational complexity (in cycles/bit) indicating how many CPU cycles are required to process 1 bit of the data \cite{9676649}.  Let $\boldsymbol{O}=\{{{o}_{{{k}_{i}}}},k\in \boldsymbol{K},i\in \boldsymbol{N}_{k}\}$ denote the offloading strategy vector. If task ${W}_{{{k}_{i}}}$ is offloaded to the surface-station via an AUV, we have ${{o}_{{{k}_{i}}}} = 1$, and ${{o}_{{{k}_{i}}}} = 0$ otherwise.  Let $\boldsymbol{r}=\{ 0 \leq {r}_{{{k}_{i}}} \leq 1,k\in \boldsymbol{K},i\in \boldsymbol{N}_{k}\}$ denote the bandwidth allocation vector to represent the specific proportion of the bandwidth resources allocated to the device ${{n}_{{{k}_{i}}}}$.  The caching strategy vector is denoted by $\boldsymbol{H}=\{{{h}_{{{k}_{i}}}},k\in \boldsymbol{K},i\in \boldsymbol{N}_{k}\}$. We have ${{h}_{{{k}_{i}}}}=1$, if the surface-station has cached the data of the task ${{W}_{{{k}_{i}}}}$ and ${{h}_{{{k}_{i}}}}=0$ otherwise. For convenience, the notations are summarized in Table \ref{NOTATIONS}.

\begin{table}[h]
\centering
\caption{NOTATIONS}
\label{NOTATIONS}
\begin{tabular}{|c|r|}
\hline
\textbf{Notation}  & \textbf{Meaning}                  \\ \hline\hline
$M$                                 & Number of AUV                                            \\ \hline
$K$                                 & Number of device group                                   \\ \hline
$f$                                 & Communication frequency                                  \\ \hline
$s$                                 & Shipping activity factor                                 \\ \hline
$w$                                 & Wind speed                                               \\ \hline
$H$                                 & Depth of water                                           \\ \hline
$\mathcal{V}$                       & Viscosity of the fluid                                   \\ \hline
$h_0$                               & Height of the IoUT device from the seabed                \\ \hline
$d_0$                               & Height of the AUV from the seabed                        \\ \hline
$r_0$                               & Radius of the vortex                                     \\ \hline
$\Omega_0$                          & Strength of the vortex                                   \\ \hline
$C_d$                               & Dragging coefficient                                         \\ \hline
$C_a$                               & Cross-sectional area                                     \\ \hline
$k_s$                               & Spreading factor                                         \\ \hline
$\rho_L$                            & Density of seawater                                      \\ \hline
$Z_{ki}$                            & Size of input data                                       \\ \hline
$f_{ki}$                            & CPU cycles per second                                    \\ \hline
$B_{\rm L}$                         & Bandwidth between AUV and device                         \\ \hline
$B_{\rm H}$                         & Bandwidth  between AUV and surface-station               \\ \hline
${P^{\rm A}_{\rm tr}}$              & Transmitted power of AUV                                \\ \hline
${P^{\rm D}_{\rm tr}}$              & Transmitted power of IoUT device                        \\ \hline
$\zeta $                            & Conversion efficiency of electricity                     \\ \hline
$\eta$                              & Overall efficiency of electronic circuitry               \\ \hline
$\Gamma_{b},\Gamma_{s}$             & Coefficient factors related to the channel gain          \\ \hline
$\omega_{k_i} $                     & Unit revenue of reducing time of IoUT device               \\ \hline
$\lambda_{k_i}$                     & Unit revenue of saving energy consumption of IoUT device \\ \hline
$\varrho $                          & Unit cost to the surface-station \\ \hline
$\chi $                             & Unit cost to the AUV      \\ \hline
\end{tabular}
\end{table}

\subsection{Communication Model}\label{B}
UAC has complex propagation characteristics, where both the multi-path effects, Doppler effects and environmental noise influence the quality of the link. For simplicity, we consider a shallow-water acoustic propagation environment assumed to be both spatially and temporally homogenous.

\subsubsection{Noise model} The environmental noise in the ocean may be caused by bubbles, shipping activity, surface wind fields, etc.  According to \cite{7006722,jensen2011computational}, the power spectral density (p.s.d) of the four main types of noise in dB per Hz at the communication frequency $f$ can be characterized by
\begin{equation}
10 \log N_{\vartheta}(f)=17-30 \log f,\label{2}
\end{equation}
\begin{equation}
10 \log N_{s}(f)=40+20\left(s-\frac{1}{2}\right)+26 \log f-60 \log (f+0.03), \label{3}
\end{equation}
\begin{equation}
10 \log N_{w}(f)=50+7.5 w^{\frac{1}{2}}+20 \log f-40 \log (f+0.4), \label{4}
\end{equation}
\begin{equation}
10 \log N_{th}(f)=-15+20 \log f, \label{5}
\end{equation}
where $N_\vartheta(f), N_{s}(f), N_{w}(f)$ and $N_{th}(f)$ represent the turbulence noise, the shipping noise, the waves noise, and the thermal noise, respectively. Furthermore, $s \in [0, 1]$ is the shipping activity factor, while $w$ represents the wind velocity  ($\mathrm{m} / \mathrm{s}$). Hence the combined noise $N(f)$ can be represented as
\begin{equation}
N(f)=N_\vartheta(f)+ N_{s}(f)+N_{w}(f)+N_{th}(f).\label{6}
\end{equation}

\begin{figure}
\centering
\subfigure[The first phase transmission.]{
\label{Firsttrans}
\includegraphics[width=8cm]{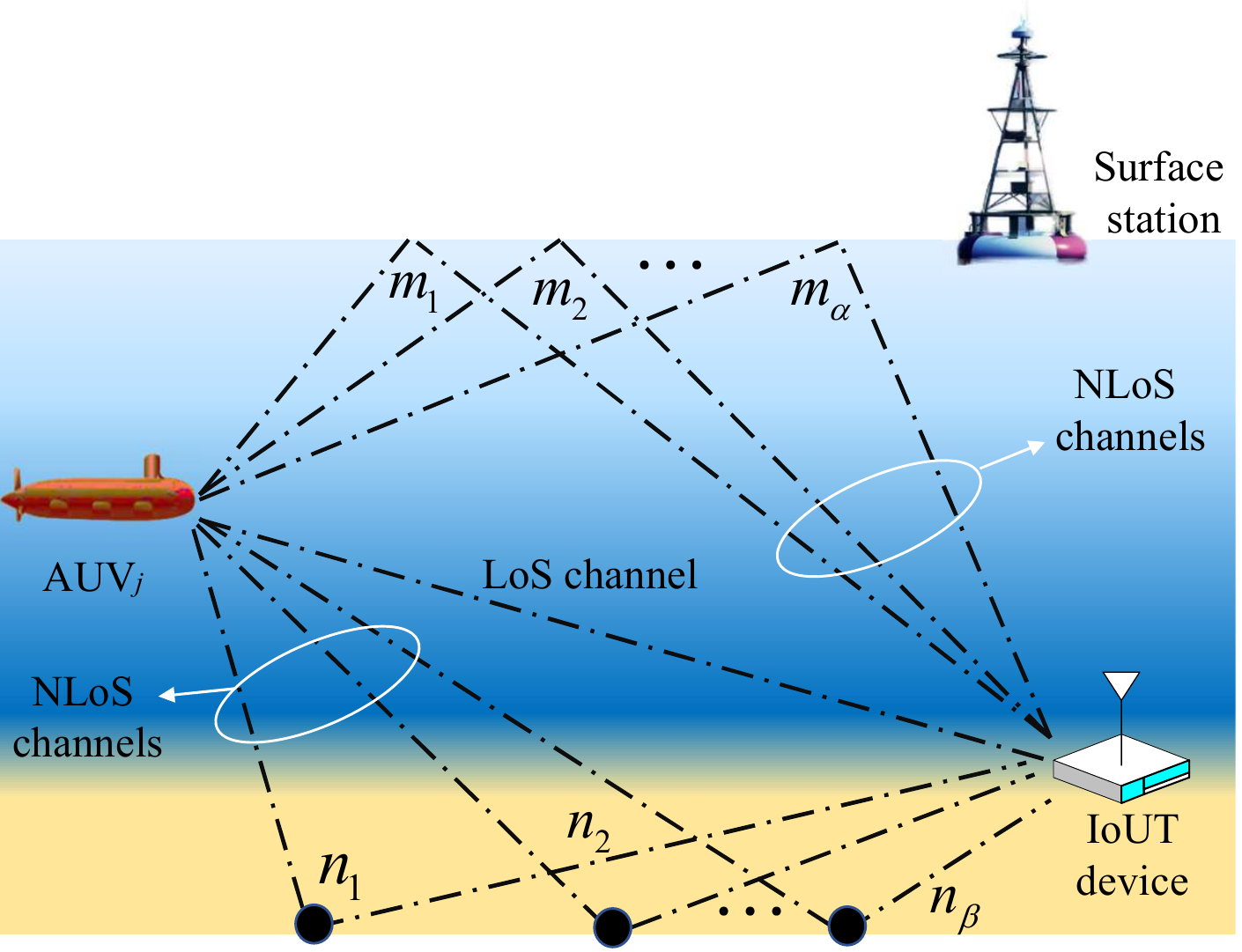}}
\subfigure[The second phase transmission.]{
\label{Secondtrans}
\includegraphics[width=8cm]{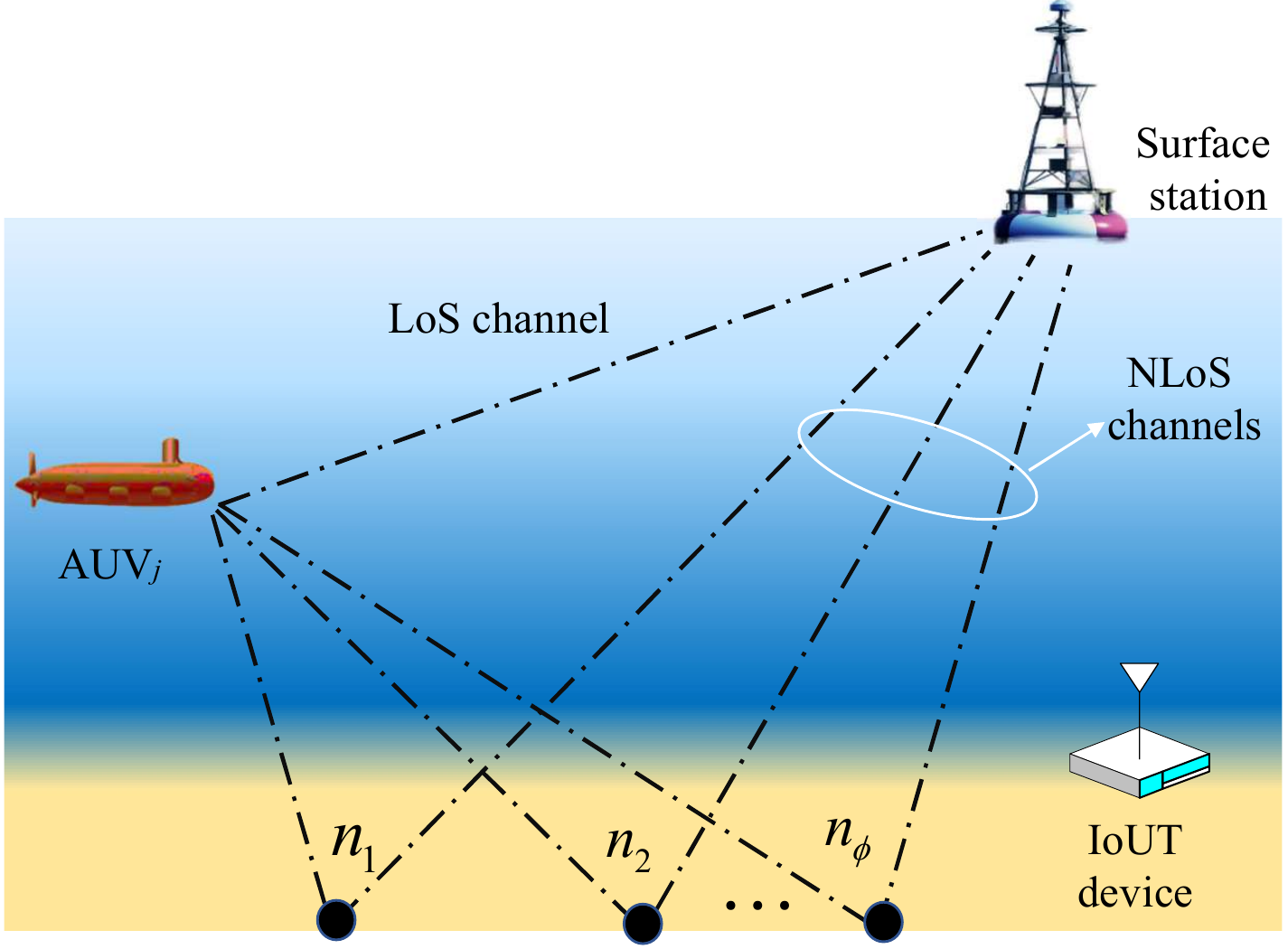}}
\caption{Multi-path effect.}
\label{MPE}
\end{figure}

There is a two-phase transmission protocol, if the IoUT devices offload their data to the surface-station, including the IoUT devices to AUV, and AUV to the surface-station phases, which can be modeled as follows:

\subsubsection{The first phase transmission: IoUT device $\to$  AUV}
The UAC channel is the superposition of the direct line-of-sight (LOS) path and a collection of non-line-of-sight (NLOS) paths, where the NLOS paths are typically reflected by underwater surfaces, the seabed and the water-air surface. Fig. \ref{Firsttrans} depicts the geometry of the UAC between IoUT devices and the AUV, where
 $\boldsymbol{m}_{u}=\left(x_{u}^{m}, y_{u}^{m}, H\right),  u \in \left\{1, 2, \dots, \alpha \right\}$ and $\boldsymbol{n}_{u}=\left(x_{u}^{n}, y_{u}^{n}, 0\right),  u \in \left\{1, 2, \dots, \beta\right\}$ are the reflection points at the sea surface and the seabed, respectively, while $H$ is the depth of water. Hence, the Euclidean distance of the LOS path is calculated as
  \begin{align}
l_{\textrm{L}}=\left\|\boldsymbol{P}_{ki}^{\textrm{S}}-\boldsymbol{P}_{j}^{\textrm{A}}\right\|_{2}, \label{LOSpath}
\end{align}
while the distance of the acoustic signal reflected from point $\boldsymbol{m}_{u}$ and point $\boldsymbol{n}_{u}$ of the NLOS propagation can be expressed as
\begin{align}
l_{m}\left(\boldsymbol{m}_{u}\right)=\left\|\boldsymbol{P}_{j}^{\textrm{A}}-\boldsymbol{m}_{u}\right\|_{2}+\left\|\boldsymbol{P}_{ki}^{\textrm{S}}-\boldsymbol{m}_{u}\right\|_{2}, \label{7}
\end{align}
and
\begin{align}
l_{n}\left(\boldsymbol{n}_{u}\right)=\left\|\boldsymbol{P}_{j}^{\textrm{A}}-\boldsymbol{n}_{u}\right\|_{2}+\left\|\boldsymbol{P}_{ki}^{\textrm{S}}-\boldsymbol{n}_{u}\right\|_{2}, \label{8}
\end{align}
respectively. Since NLOS paths have lost much of their energy after multiple reflections, we only have to pay attention to a finite number of significant paths  \cite{7006722}.  Furthermore, obtaining the lower bound of the signal-to-noise ratio (SNR) is more beneficial by finding the minimum NLOS path lengths. We can easily to calculate the shortest NLOS path lengths $l_{m}\left(\boldsymbol{m}^{\star}\right)$ and $l_{n}\left(\boldsymbol{n}^{\star}\right)$  reflected from the top and bottom surfaces as
\small
\begin{align}
l_{m}\left(\boldsymbol{m}^{\star}\right) = \sqrt{\left(x_{j}^{\textrm{A}}-x_{ki}^{\textrm{S}}\right)^{2}+\left(y_{j}^{\textrm{A}}-y_{ki}^{\textrm{S}}\right)^{2}+\left(2 {H}-h_{0}-d_{0}\right)^{2}}, \label{9}
\end{align}
and
\begin{align}
l_{n}\left(\boldsymbol{n}^{\star}\right) = \sqrt{\left(x_{j}^{\textrm{A}}-x_{ki}^{\textrm{S}}\right)^{2}+\left(y_{j}^{\textrm{A}}-y_{ki}^{\textrm{S}}\right)^{2}+\left(h_{0}+d_{0}\right)^{2}} , \label{10}
\end{align}
\normalsize
respectively.

Let $A(l, f)$ be the attenuation at frequency $f$ over the distance $l$, which is given by
\begin{equation}\label{11}
A(l, f)=l^{k_s} a(f)^{l},
\end{equation}
where $k_s$ represents the spreading factor, and $a(f)$ is the absorption coefficient, which can be expressed empirically in $\mathrm{dB}$ per $\mathrm{km}$ with $f$ in KHz as follows \cite{9072166}
\begin{equation}\label{12}
10 \log a(f)=\frac{0.11 f^{2}}{1+f^{2}}+\frac{44 f^{2}}{4100+f^{2}}+2.75 \cdot 10^{-4} f^{2}+0.003.
\end{equation}
Therefore, the normalized SNR of a signal with unity transmitted power and bandwidth can be represented as
\begin{equation}\label{13}
\gamma(l, f)=  \frac{1}{A(l, f) N(f)}.
\end{equation}
The lower bound of the SNR considering the minimum NLOS path lengths derived from Eq. (\ref{9}) and Eq. (\ref{10})  can be expressed as \cite{abichandani2015mixed}
\small
\begin{eqnarray}
\label{14}
\begin{aligned}
&\gamma\left(l_{\textrm{L}}, f\right)_{\min }=\frac{1}{N(f)}\cdot\{\frac{1}{\sqrt{A\left(l_{\textrm{L}}, f\right)}}-\frac{\alpha \Gamma_{s}}{\sqrt{A\left(l_{m}\left(\boldsymbol{m}^{\star}\right), f\right)}} \\
&\left.-\frac{\beta \Gamma_{b}}{\sqrt{A\left(l_{n}\left(\boldsymbol{n}^{\star}\right), f\right)}}\right\}^{2},
\end{aligned}
\end{eqnarray}
\normalsize
where $\Gamma_{s}$ and $\Gamma_{b}$ characterize the channel gain of the shortest NLOS path reflected from the top and bottom surfaces, respectively.

Hence, the data rate between the  $\text{AUV}_j$ and IoUT device ${{n}_{{{k}_{i}}}}$ can be formulated as
\begin{equation}
R_{{{k}_{i}}}^{\textrm{DA}}={{r}_{{{k}_{i}}}}B_{\textrm{L}} \log _{2}\left(1+\frac{\eta P^{\textrm{D}}_{\textrm{tr}} \gamma\left(l_{\textrm{L}}, f\right)_{\min }}{2 \pi H_1 (1 \mu \mathrm{Pa}) {{r}_{{{k}_{i}}}}B_{\textrm{L}}}\right) \label{15},
\end{equation}
where $B_{\textrm{L}}$ is the total bandwidth of $\text{AUV}_j$, while $P^{\textrm{D}}_{\textrm{tr}}$ denotes the transmitted power of the IoUT device, respectively. Furthermore, $\eta$ is the overall efficiency of the electronic circuitry  including both the power amplifier and transducer \cite{duan2020value}, while $H_1$ is the water depth of IoUT devices and ${{r}_{{{k}_{i}}}}$ denotes the proportion of the bandwidth allocated to the IoUT device ${{n}_{{{k}_{i}}}}$, which satisfies
\begin{eqnarray}
\label{16}
\begin{cases}
{{r}_{{{k}_{i}}}}=0, \: if \: {{o}_{{{k}_{i}}}}=0, \forall k\in \boldsymbol{K},i\in \boldsymbol{N}_k,\\
{{r}_{{{k}_{i}}}}=0, \: if \: {{h}_{{{k}_{i}}}}=1,\forall k\in \boldsymbol{K},i\in \boldsymbol{N}_k,\\
\sum\limits_{i=1}^{{{N}_{k}}}{{o}_{{{k}_{i}}}}({1-{h}_{{{k}_{i}}}}){{{r}_{{{k}_{i}}}}\le 1}, \forall k\in \boldsymbol{K}.
\end{cases}
\end{eqnarray}

\subsubsection{The second phase transmission: AUV  $\to$ surface-station}
As shown in Fig. \ref{Secondtrans},  $\boldsymbol{w}_{u}=\left(x_{u}^{w}, y_{u}^{w}, 0\right),  u \in \left\{1, 2, \dots, \phi \right\}$ is the reflection point at the seabed of multi-path propagation. Hence, the Euclidean distance of the LOS path and the NLOS path is given by
\begin{align}
l_{\textrm{H}}=\left\|{\boldsymbol{P}}_{j}^{\textrm{A}}-{\boldsymbol{P}}_{\textrm{SS}}\right\|_{2} \label{17}
\end{align}
and
\begin{align}
l_{{w}}^{\textrm{H}}\left({{\boldsymbol{w}}_{u}}\right)=\left\|{\boldsymbol{P}}_{j}^{\textrm{A}}-{{\boldsymbol{w}}_{u}}\right\|_{2}+\left\|{{\boldsymbol{w}}_{u}}-{\boldsymbol{P}}_{\textrm{SS}}\right\|_{2}, \label{18}
\end{align}
respectively. Thus we can obtain the minimum NLOS path as
\begin{align}
l_{w}^{\textrm{H}}\left(\boldsymbol{w}^{\star}\right)=\sqrt{\left(x_{j}^{\textrm{A}}\right)^{2}+\left(y_{j}^{\textrm{A}}\right)^{2}+\left(h_{0}+d_{0}\right)^{2}}. \label{19}
\end{align}
Similar to Eq. (\ref{11})-(\ref{14}),  the lower bound of SNR $\gamma\left(l_{\textrm{H}}\right)_{\min }$ at the surface-station subjected to NLOS propagation is given by
\small
\begin{eqnarray}
\label{22}
\begin{aligned}
\gamma\left(l_{\textrm{H}}, f\right)_{\min }=\frac{1}{N(f)}\cdot\{\frac{1}{\sqrt{A\left(l_{\textrm{H}}, f\right)}}-\left.\frac{\beta \Gamma_{b}}{\sqrt{A\left(l_{w}^\textrm{H}\left(\boldsymbol{w}^{\star}\right), f\right)}}\right\}^{2},
\end{aligned}
\end{eqnarray}
\normalsize

The classic code division multiple access (CDMA) is adopted for the UAC links between the AUVs and surface-station, and the data rate between them can be calculated as
\begin{eqnarray}
\label{23}
R_{k}^{\textrm{AS}}=B_{\textrm{H}} \log _{2}\left(1+\frac{\eta P^{\textrm{A}}_{\textrm{tr}} \gamma\left(l_{\textrm{H}}, f\right)_{\min }}{2 \pi H_2 (1 \mu \mathrm{Pa})B_{\textrm{H}}}\right),
\end{eqnarray}
where $\boldsymbol{P}^{\textrm{A}}_{\textrm{tr}}$ represents the transmitted power of the AUV, while $B_{\textrm{H}}$ and $H_2$ represent the  available bandwidth and the water depth of AUVs, respectively.

\subsection{Caching Model}\label{C}
Because there are often repeated requests for tackling the same task, caching some data of the previous requested task is capable of reducing the backhaul latency and alleviate the pressure on the backhaul bandwidth \cite{8305608,yang2019joint}. If task ${{W}_{{{k}_{i}}}}$ is cached by the surface-station, ${{h}_{{{k}_{i}}}}=1$  and ${{o}_{{{k}_{i}}}}$ should be 1,  to avoid it being processed locally on the device for saving device's energy and reducing the processing latency. Hence the binary variable ${{h}_{{{k}_{i}}}}$ of caching decision should satisfy
\begin{eqnarray}
\label{31}
{{h}_{{{k}_{i}}}}\le {{o}_{{{k}_{i}}}}.
\end{eqnarray}
Moreover, since the storage capacity of surface-station is typically limited, the caching strategy should satisfy \cite{9090328}
\begin{eqnarray}
\label{32}
\sum\limits_{k=1}^{K}{\sum\limits_{i=1}^{{{N}_{k}}}{{h_{{k}_{i}}}{{Z}_{{{k}_{i}}}}}}\le {{C}_{e}},
\end{eqnarray}
where $C_e$ is the maximal storage capacity of the surface-station.

\subsection{Computing Model}\label{D}
Next we discuss the processing time of local vs. offloaded surface-station based computation.

\subsubsection{ Local Computing} The computing capability of ${{n}_{{{k}_{i}}}}$ is denoted by ${{f}_{{{k}_{i}}}}$ and different IoUT devices have different computing capabilities. The duration of completing the task ${{W}_{{{k}_{i}}}}$ locally is calculated as
\begin{eqnarray}
\label{24}
T_{{{k}_{i}}}^{\textrm{L}}=\frac{{{\alpha}_{{{k}_{i}}}}{{Z}_{{{k}_{i}}}}}{{{f}_{{{k}_{i}}}}}.
\end{eqnarray}

\subsubsection{Computing at surface-station} If  task ${{W}_{{{k}_{i}}}}$ is offloaded to the surface-station for processing, two-stage transmission is needed. But if  task ${{W}_{{{k}_{i}}}}$ has already been cached at the surface-station, the data transmission procedure is eliminated. The transmission time of tackling the task at the surface-station can be represented as
\begin{eqnarray}
\label{25}
T_{{{k}_{i}}}^{\textrm{T}}=T_{{{k}_{i}}}^{\textrm{DA}}+T_{{{k}_{i}}}^{\textrm{AS}},
\end{eqnarray}
 where $T_{{{k}_{i}}}^{\textrm{DA}}$ and $T_{{{k}_{i}}}^{\textrm{AS}}$ is the duration of transmitting the data from the IoUT device to the AUV and that from the AUV to the surface-station, respectively. The duration of downloading the results from the surface-station is usually ignored, since the results are more compact than the input data size of ${{W}_{{{k}_{i}}}}$ \cite{6787113}. To elaborate, $T_{{{k}_{i}}}^{\textrm{DA}}$ and $T_{{{k}_{i}}}^{\textrm{AS}}$ are calculated as
 \begin{equation}
\label{25-1}
T_{{{k}_{i}}}^{\textrm{DA}}=(1-{{h}_{{{k}_{i}}}})\frac{{{Z}_{{{k}_{i}}}}}{R_{{{k}_{i}}}^{\textrm{DA}}},
\end{equation}
and
\begin{eqnarray}
\label{25-2}
T_{{{k}_{i}}}^{\textrm{AS}}=(1-{{h}_{{{k}_{i}}}})\frac{{{Z}_{{{k}_{i}}}}}{R_{k}^{\textrm{AS}}},
\end{eqnarray}
 respectively.

Furthermore, to improve efficiency and reduce the energy consumption, it is beneficial to sparingly activate the limited computing resources. Assume that the total computational resource allocated for each AUV by surface-station is denoted by $F$,  and $\boldsymbol{F}=\{{{f}_{{{k}_{i}}}^m} ,k\in \boldsymbol{K},i\in \boldsymbol{N}_{k}\}$ represents the computing resource allocation vector, where $f_{{{k}_{i}}}^{m} \in [0,1]$ is the specific proportion of $F$  allocated to ${{n}_{{{k}_{i}}}}$.  Hence the computing time of task ${{W}_{{{k}_{i}}}}$ at the surface-station is given by
\begin{eqnarray}
\label{26}
T_{{{k}_{i}}}^{\textrm{M}}=\frac{{{\alpha}_{{{k}_{i}}}}{{Z}_{{{k}_{i}}}}}{f_{{{k}_{i}}}^{m}F},
\end{eqnarray}
where the computing resource allocation vector should satisfy
\begin{eqnarray}
\label{27}
\begin{cases}
{{f}_{{{k}_{i}}}^m}=0, \: if \: {{o}_{{{k}_{i}}}}=0, \forall k\in \boldsymbol{K},i\in \boldsymbol{N}_k,\\
\sum\limits_{i=1}^{{{N}_{k}}}{{o}_{{{k}_{i}}}}{{{f}_{{{k}_{i}}}^m}\le 1}, \forall k\in \boldsymbol{K}.
\end{cases}
\end{eqnarray}

Overall, the total duration of tackling task ${{W}_{{{k}_{i}}}}$ is represented as
\begin{eqnarray}
\label{28}
{{T}_{{{k}_{i}}}}={o_{{k}_{i}}}(T_{{{k}_{i}}}^{M}+T_{{{k}_{i}}}^{T})+(1-{o_{{k}_{i}}})T_{{{k}_{i}}}^{L},
\end{eqnarray}
while that of addressing all tasks in $DG_k$ is given by
\begin{eqnarray}
\label{29}
{{A}_{{{k}}}}=\max \limits_{i=\left\{1,2,\dots,{N}_{k}\right\}} {{T}_{{k}_{i}}}, \forall k\in \boldsymbol{K}.
\end{eqnarray}

\subsection{Trajectory Model}\label{E}
We assume that $AUV_j$ servers $S_j$ DGs, and we define $\boldsymbol{P}_{j}^{\textrm{A}}[\xi], \forall \xi \in \boldsymbol{S_j}=\{0,1,2,\ldots,S_j,S_j+1\}$ as the trajectory of $AUV_j$, which starts from the surface-station, i.e., $\boldsymbol{P}_{j}^{\textrm{A}}[0]$, passes through all the assigned DGs and finally returns to the surface-station, i.e., $\boldsymbol{P}_{j}^{\textrm{A}}[S_j+1]$ for recharging. Therefore, we have
\begin{eqnarray}
\label{33}
\boldsymbol{P}_{j}^{\textrm{A}}[S_j+1]=\boldsymbol{P}_{j}^{\textrm{A}}[0], \forall j \in \boldsymbol{M}.
\end{eqnarray}
We define $d_j[\xi]$ as the distance between the two points, which can be calculated as
\begin{equation}\small
\label{34}
{{d}_{j}}[\xi]=\left\|{{\boldsymbol{P}}_{j}}^{\textrm{A}}[\xi+1]-{{\boldsymbol{P}}_{j}}^{\textrm{A}}[\xi]\right\|_2, \forall j \in \boldsymbol{M}, \forall \xi \in \{0,1,2,\ldots,S_j\}.
\end{equation}
Let ${{Y}_{{{j}_{k}}}}[\xi]=1$ represent that $AUV_j$ selects the  $DG_k$ as its $\xi$-th hovering DG, otherwise ${{Y}_{{{j}_{k}}}}[\xi]=0$. The AUV trajectory design strategy can be represented by $\boldsymbol{Y}=\{{{Y}_{{{j}_{k}}}}[\xi], j \in \boldsymbol{M}, \xi \in \boldsymbol{S_j}, k \in \boldsymbol{K}$\}. In order to guarantee that each DG can be covered and served only once, we have
\begin{eqnarray}
\label{35}
\sum\limits_{j=1}^{M}{\sum\limits_{\xi=1}^{{{S}_{j}}}{{{Y}_{{{j}_{k}}}}[\xi]}}=1, \forall j \in \boldsymbol{M},  \forall k \in \boldsymbol{K},
\end{eqnarray}
\begin{eqnarray}\label{36}
\sum\limits_{\xi=1}^{{{S}_{j}}}{\sum\limits_{k=1}^{K}{{{Y}_{{{j}_{k}}}}}}[\xi]={{S}_{j}}, \forall j \in \boldsymbol{M},
\end{eqnarray}
and
\begin{eqnarray}
\label{37}
&\sum\limits_{j=1}^{M}{{{S}_{j}}}=K, \forall j \in \boldsymbol{M}.
\end{eqnarray}
Therefore, the total hovering time of $AUV_j$  can be expressed as
\begin{eqnarray}
\label{38}
T_{j}^{\rm H}=\sum\limits_{k=1}^{K}{\sum\limits_{\xi=1}^{{{S}_{j}}}{{{Y}_{{{j}_{k}}}}[\xi]{{A}_{k}}}},
\end{eqnarray}
where trajectory of $AUV_j$ is composed of $S_j+1$ sub-trajectories. We assume that each AUV moves along the segment at a constant velocity $V_{k}$. Therefore, the time of $AUV_j$ in each sub-trajectory is given by
\begin{eqnarray}
\label{40}
t_{j}^{\textrm{F}}[\xi]=\frac{{{{d}_{j}}[\xi]}}{V_{k}},  \forall j \in \boldsymbol{M}, \forall \xi \in \{0,1,2,\ldots,S_j\}.
\end{eqnarray}
Furthermore, the total travelling distance and the travelling time of $AUV_j$ are
\begin{eqnarray}
\label{39}
{{I}_{j}}=\sum\limits_{\xi=0}^{{{S}_{j}}}{{{d}_{j}}[\xi]},\forall j \in \boldsymbol{M},
\end{eqnarray}
and
\begin{eqnarray}
\label{40}
T_{j}^{\textrm{F}}=\frac{{{I}_{j}}}{V_{k}},
\end{eqnarray}
respectively. Consequently, the total cruising time of $AUV_j$ in a cycle is given by
\begin{eqnarray}
\label{41}
{{T}_{j}^{\textrm{AT}}}=T_{j}^{\textrm{F}}+T_{j}^{\textrm{H}},\forall j \in \boldsymbol{M}.
\end{eqnarray}
To strike a balance, we define $\varepsilon$ as a constraint for limiting the difference of travelling time among different AUVs as
\begin{eqnarray}
\label{42}
{{T}^{\textrm{AT}}_{\max }}-{{T}^{\textrm{AT}}_{\min }}\le \varepsilon,
\end{eqnarray}
where ${{T}^{\textrm{AT}}_{\max }} = \max \left\{ {{T}_{j}^{\textrm{AT}}} ,  j \in \boldsymbol{M}  \right\}$ and ${{T}^{\textrm{AT}}_{\min }} = \min \left\{ {{T}_{j}^{\textrm{AT}}} ,  j \in \boldsymbol{M}  \right\}$.
\subsection{Motion Model}\label{F}
The underwater oceanic environment is complex and hostile with dynamically fluctuating water velocity, vortex, etc., which may impose a significant impact on the AUV's movement. To quantify it, we construct a model for evaluating the effects of the turbulent oceanic environments on AUV's motion based on  the Navier-Stokes equation \footnote{In practice, most commercial AUVs are equipped with the horizontal acoustic Doppler current profiler (H-ADCP) and Doppler velocity logger (DVL), which can measure ocean current velocity profiles up to hundreds of meters in front of the AUV with an accuracy of 1\% of the measured magnitude $\pm$ 5 mm/s \cite{zeng2015efficient}.  }. Specifically, the oceanic current field can be represented as \cite{9612189}
\begin{equation}\label{43}
\frac{\partial \boldsymbol \omega}{\partial t}+({\vec{V_{\rm C}}} \nabla) \boldsymbol \omega=\mathcal{V} \Delta \boldsymbol\omega,
\end{equation}
where ${\vec{V_{\rm C}}}=\left(V_{x}, V_{y},V_{z}\right)$ represents the velocity field, while $\boldsymbol\omega = (\frac{\partial V_{z}}{\partial y}-\frac{\partial V_{y}}{\partial z})\vec{i}+(\frac{\partial V_{x}}{\partial z}-\frac{\partial V_{z}}{\partial x})\vec{j}+(\frac{\partial V_{y}}{\partial x}-\frac{\partial V_{x}}{\partial y})\vec{k}$  denotes the vorticity of the current. Furthermore, $\mathcal{V}$ is the viscosity of the fluid, while $\nabla$ and $\Delta$ represent the gradient and Laplacian operator, respectively.
To facilitate the analysis, we approximate the Navier-Stokes equation as
\begin{equation}\label{44}
V_{x}(\boldsymbol{P}_{j}^{\textrm{A}})=-\frac{\Omega_{0} \cdot\left(y_{j}^{\textrm{A}}-y_{0}\right)}{2 \pi\left\|\boldsymbol{P}_{j}^{\textrm{A}}-\boldsymbol{P}_{0}\right\|_{2}^{2}} \left(1-e^{-\frac{\left\|\boldsymbol{P}_{j}^{\textrm{A}}-\boldsymbol{P}_{0}\right\|_{2}^{2}}{r_{0}^{2}}}\right),
\end{equation}
\begin{equation}\label{45}
V_{y}(\boldsymbol{P}_{j}^{\textrm{A}})=\frac{\Omega_{0} \cdot\left(x_{j}^{\textrm{A}}-x_{0}\right)}{2 \pi\left\|\boldsymbol{P}_{j}^{\textrm{A}}-\boldsymbol{P}_{0}\right\|_{2}^{2}} \left(1-e^{-\frac{\left\|\boldsymbol{P}_{j}^{\textrm{A}}-\boldsymbol{P}_{0}\right\|_{2}^{2}}{r_{0}^{2}}}\right),
\end{equation}
\begin{equation}\label{46}
V_{z}(\boldsymbol{P}_{j}^{\textrm{A}})=\frac{\Omega_{0} \cdot\left(d_{0}-z_{0}\right)}{2 \pi\left\|\boldsymbol{P}_{j}^{\textrm{A}}-\boldsymbol{P}_{0}\right\|_{2}^{2}} \left(1-e^{-\frac{\left\|\boldsymbol{P}_{j}^{\textrm{A}}-\boldsymbol{P}_{0}\right\|_{2}^{2}}{r_{0}^{2}}}\right),
\end{equation}
and
\begin{equation}\label{47}
\boldsymbol\omega(\boldsymbol{P}_{j}^{\textrm{A}})=\frac{\Omega_{0}}{\pi r_{0}^{2}} e^{-\frac{\left\|\boldsymbol{P}_{j}^{\textrm{A}}-\boldsymbol P_{0}\right\|_{2}^{2}}{r_{0}^{2}}},
\end{equation}
where $\boldsymbol{P}_{j}^{\textrm{A}}$ and $\boldsymbol{P}_{0}=\left(x_{0}, y_{0}, z_{0}\right)$ denote the coordinates of $AUV_j$ and the center of the Lamb vortex  \cite{shuai2022accelerated}, respectively. Furthermore $\Omega_{0}$ and $r_{0}$ represent the strength and radius of the vortex, respectively. In fact, most of the energy consumption of the AUV is dissipated by overcoming the resistance of the water for maintaining the velocity $V_{k}$. To determine the propulsion force of $AUV_j$ required for maintaining a given velocity $V_{k}$, the relative velocity between the  AUV and the current should be derived, which can be expressed as
\begin{equation}\label{48}
\vec{V}_{R_{k}}(\boldsymbol{P}_{j}^{\textrm{A}})=V_{k} \cdot \vec{e}_{k}-\vec{V}_{\rm C}(\boldsymbol{P}_{j}^{\textrm{A}}),
\end{equation}
where  $\vec{V}_{{\rm C}}(\boldsymbol{P}_{j}^{\textrm{A}})$ denotes the water flow velocity, while $ \vec{e}_{k}$ is the unit vector of the direction of the $AUV_j$. According to classic computational fluid dynamics (CFD) methods \cite{bhatti2020recent},  the drag force required for floating and for moving can be expressed as
\begin{equation}\label{49-a}
F_{j}^{\rm H} =  \frac{1}{2} \rho_{\rm L}\left\|{\vec{V}_{\rm C} (\boldsymbol{P}_{j}^{\textrm{A}})}\right\|_{2}^{2}C_a C_{d},
\end{equation}
and
\begin{equation}\label{49}
F_{j}^{\rm F} =  \frac{1}{2} \rho_{\rm L}\left\|{\vec{V}_{R_{k}} (\boldsymbol{P}_{j}^{\textrm{A}})}\right\|_{2}^{2} C_a  C_{d},
\end{equation}
respectively, where $C_{d}$ denotes the dragging coefficient, while $\rho_{\rm L}$ and $C_a$ represent the density of seawater and the cross-sectional area of the AUV moving along the current direction.
\subsection{Energy Consumption Model}\label{G}
In the following, we analyze the energy consumption of the MTUC from the perspective of the user (i.e., IoUT devices) and the service provider (i.e., surface-station and AUVs), respectively.

\subsubsection{The energy consumption of users}
For IoUT devices, the energy is mainly consumed either by local computations or by transmissions, when tasks are offloaded to the surface-station. If task ${{W}_{{{k}_{i}}}}$ solved locally, the energy consumption of computing is formulated by
\begin{eqnarray}
\label{50}
E_{{{k}_{i}}}^{\rm L}=\mu {{({{f}_{{{k}_{i}}}})}^{\sigma }}T_{{{k}_{i}}}^{\rm L},
\end{eqnarray}
where ${{f}_{{{k}_{i}}}}$ is the CPU frequency of the IoUT device ${{n}_{{{k}_{i}}}}$. According to \cite{9301339,Dinh2017}, $\mu$ is a constant that depends on the average switched capacitance and the average activity factor, while $\sigma$ is a constant close to 3. By contrast, if task ${{W}_{{{k}_{i}}}}$ is transmitted to the AUV for further processing at the surface-station, the corresponding transmit energy consumption consumed of  IoUT device ${{n}_{{{k}_{i}}}}$ is given by \cite{9451536}
\begin{equation}\label{51}
E_{{{k}_{i}}}^{\rm DA}=(1-{{h}_{{{k}_{i}}}})\frac{2 \pi (1 \mu \mathrm{Pa})  {{r}_{{{k}_{i}}}}B_{\textrm{L}}}{\eta \gamma\left(l_{L}, f\right)_{\min }} \left[2^{\frac{R_{{{k}_{i}}}^{\rm DA}}{{{r}_{{{k}_{i}}}}B_{\textrm{L}} {{T}_{{{k}_{i}}}^{\rm DA}}}}-1\right] {T_{{{k}_{i}}}^{\rm DA}}.
\end{equation}

\subsubsection{The energy consumption of service provider}
The energy consumption of the service provider is composed of two parts, including that of the surface-station solving the tasks and that of the AUVs for cruising and forwarding the tasks.

Specifically, for the surface-station, similar to \eqref{50}, when task ${{W}_{{{k}_{i}}}}$ is offloaded to the surface-station, the energy consumption of this is given by
\begin{eqnarray}
\label{52}
E_{{{k}_{i}}}^{\rm M}=\mu {{(f_{{{k}_{i}}}^{m}F)}^{\sigma }}T_{{{k}_{i}}}^{\rm M}.
\end{eqnarray}
Furthermore, as for the AUVs, similar to Eq. (\ref{51}), upon forwarding task ${{W}_{{{k}_{i}}}}$ from ${{n}_{{{k}_{i}}}}$ to a surface-station, the transmit energy consumption consumed by a AUV is formulated by
\begin{equation}\label{53}
E_{{{k}_{i}}}^{\rm AS}=(1-{{h}_{{{k}_{i}}}})\frac{2 \pi (1 \mu \mathrm{Pa})B_{\rm{H}}}{\eta \gamma\left(l_{\rm H}, f\right)_{\min }} \left[2^{\frac{R_{{k}}^{\rm AS}}{B_{\rm {H}} \cdot {{T_{{{k}_{i}}}^{\rm AS}}}}}-1\right]  {T_{{{k}_{i}}}^{\rm AS}}.
\end{equation}
As for the energy consumption of the AUV's movement, it should be discussed in two scenarios, namely for hovering above the DGs and for moving between two destinations.  According to Eq. (\ref{49-a}), the drag force required to stay afloat above the $\xi$-th DG is formulated by
\begin{equation}\label{58}
F_{j}^{\rm H}[\xi]=\frac{1}{2} \rho_{\rm L}\left\|{\vec{V}_C(\boldsymbol{P}_{j}^{\textrm{A}}[\xi])}\right\|_{2}^{2} C_a C_{d}.
\end{equation}
Consequently, the electric power generating the required force is calculated as
\begin{eqnarray}
\label{ElectricPowerForSuspending}
\begin{aligned}
{{P}^{\rm H}_j}[\xi] =  \frac{1}{\zeta} F_{j}^{\rm H}[\xi] \left\|{\vec{V}_C}(\boldsymbol{P}_{j}^{\textrm{A}}[\xi])\right\|_{2},
\end{aligned}
\end{eqnarray}
where $\zeta$ is the electricity conversion efficiency. Since the water flow velocity is different at each point during the movement of the AUV, this will impose significant challenges on our further analysis. Therefore, we approximate the average relative flow velocity in a  sub-trajectory by the average of the relative flow velocity at the starting point, the midpoint and the end of this sub-trajectory. The more DGs are deployed in the same area, the closer the approximation to reality. The average relative flow velocity of $AUV_j$ moving from the $\xi$-th DG to the $(\xi+1)$-st DG is given by
\begin{eqnarray}
\label{AverageRelativeFlowVelocity}
\begin{aligned}
&\overline{\vec{V}_{R_{k}}(\boldsymbol{P}_{j}^{\textrm{A}}[\xi])}=\\
&\frac{1}{3}(\vec{V}_{R_{k}}(\boldsymbol{P}_{j}^{\textrm{A}}[\xi])+\vec{V}_{R_{k}}(\boldsymbol{P}_{j}^{\textrm{A}}[\xi+1])+\vec{V}_{R_{k}}(\hat{\boldsymbol{P}_{j}^{\textrm{A}}}[\xi_{mid}]))\\
& \forall \xi \in \{0,1,2,\ldots,S_j\},
\end{aligned}
\end{eqnarray}
where $\boldsymbol{P}_{j}^{\textrm{A}}[\xi]$ and $\boldsymbol{P}_{j}^{\textrm{A}}[\xi+1]$ are the coordinates of the $\xi$-th DG and the $(\xi+1)$-st DG, while  $\hat{\boldsymbol{P}_{j}^{\textrm{A}}}[\xi_{mid}]$ represents the coordinates of the middle point between the $\xi$-th DG and the $(\xi+1)$-st DG.  Therefore, according to Eq. (\ref{49}), the drag force required for supporting $AUV_j$ movement from the $\xi$-th DG to the $(\xi+1)$-st DG is calculated as
\begin{equation}\label{PullingForceForMovingFromASubjectory}
F_{j}^{\rm F}[\xi]=\frac{1}{2} \rho_{L}\left\|\overline{\vec{V}_{R_{k}}(\boldsymbol{P}_{j}^{\textrm{A}}[\xi])}\right\|_{2}^{2}C_a C_{d}.
\end{equation}
Consequently, the corresponding electric power is represented by
\begin{eqnarray}
\label{ElectricPowerForMovingFromASubjectory}
\begin{aligned}
{{P}^{\rm F}_j}[\xi] =  \frac{1}{\zeta} \cdot F_{j}^{\rm F}[\xi] \cdot\left\|\overline{{\vec{V}_{R_{k}}}(\boldsymbol{P}_{j}^{\textrm{A}}[\xi])}\right\|_{2}.
\end{aligned}
\end{eqnarray}
As a result, the energy consumption of $AUV_j$ cruising through a specific cycle is given by
\begin{eqnarray}
\label{38}
E_{j}=\sum\limits_{k=1}^{K}{\sum\limits_{\xi=1}^{{{S}_{j}}}{{{Y}_{{{j}_{k}}}}[\xi]{{A}_{k}}}P_j^{\rm H}[\xi] }+{\sum\limits_{\xi=0}^{{{S}_{j}}}{{{t}^{\rm F}_{{{j}}}}[\xi]}P_j^{\rm F}[\xi]  }.
\end{eqnarray}

\subsection{Utility Function}\label{H}
Our proposed MTUC framework aims for maximizing the profit of the whole system. Specifically, the latency and energy consumption improvement of the users, i.e., IoUT devices, are deemed to be the revenue, while the cost is the energy consumption imposed on the service provider, namely the AUV and the surface-station. The profit is calculated by the revenue minus cost.

To elaborate, with the assistance of our MTUC framework, the computation latency and energy consumption of the task ${W}_{{{k}_{i}}}$ can be reduced to
\begin{eqnarray}
\label{56}
T_{{{k}_{i}}}^{\rm S}={{o}_{{{k}_{i}}}}\left[T_{{{k}_{i}}}^{\rm L}-(T_{{{k}_{i}}}^{\rm M}+ T_{{{k}_{i}}}^{\rm T})\right],
\end{eqnarray}
and
\begin{eqnarray}\label{57}
E_{{{k}_{i}}}^{\rm S}={{o}_{{{k}_{i}}}}(E_{{{k}_{i}}}^{\rm L}-E_{{{k}_{i}}}^{\rm DA}),
\end{eqnarray}
respectively.
Accordingly, the revenue that the MTUC framework can obtain is given by \cite{chenmeng2017,9322229}
\begin{eqnarray}
\label{58}
Re =\sum\limits_{k=1}^{K}{\sum\limits_{i=1}^{{{N}_{k}}}{{{\omega}_{{{k}_{i}}}T_{{{k}_{i}}}^{\rm S}} +{{\lambda}_{{{k}_{i}}}E_{{{k}_{i}}}^{\rm S}}}},
\end{eqnarray}
where the ${\omega}_{{{k}_{i}}}$ and ${\lambda}_{{{k}_{i}}}$ are the unit revenue attained by reducing the time and by saving energy for the IoUT device ${n}_{{{k}_{i}}}$, respectively.

The cost that the MTUC framework has to bear is composed of the cost of solving the computing task and supporting the AUVs' movements. Specifically,  the cost of the surface-station and of the AUV for solving task ${{W}_{{{k}_{i}}}}$ is given by
\begin{eqnarray}
\label{59_1}
CT^{\rm M}_{{k}_{i}}={{o}_{{{k}_{i}}}}\varrho E_{{{k}_{i}}}^{\rm M},
\end{eqnarray}
and
\begin{eqnarray}
\label{59_2}
CT^{\rm AS}_{{k}_{i}}={{o}_{{{k}_{i}}}}\chi E_{{{k}_{i}}}^{\rm AS},
\end{eqnarray}
respectively, where ${\varrho}$ denotes the unit cost to the surface-station, while ${\chi}$ is the unit cost to the AUV. It is noted that the values  of ${\varrho}$ and ${\chi}$ are different because of the difference in the difficulty of replenishing the energy of the surface-station and of the AUV.  Therefore, the cost of solving all the tasks for the MTUC framework is calculated as
\begin{eqnarray}
\label{60}
CT= \sum\limits_{k=1}^{K}{\sum\limits_{i=1}^{{{N}_{k}}}\left(CT^{\rm M}_{{k}_{i}}+  CT^{\rm AS}_{{k}_{i}}  \right)},
\end{eqnarray}
In fact, most of the energy consumption is dissipated by the AUV's movement\footnote{A DG typically has a dozen to dozens of IoUT devices \cite{9451536,duan2020value,9072166}. Assume that each device has an image processing task that has to be solved with the data size of 300 Kb and computational complexity of 2000 cycles/bit. If all the IoUT devices select to offload their task to surface-station, the transmit energy consumption of the AUV is about 5000 J, while the computing energy consumption of the surface-station is about 1000 J. Furthermore, the energy consumption of the AUV for floating above a DG and for moving from the DG to the next DG is about 6000 J. Hence it is feasible to consider the energy consumption of computation and transmission together with the energy consumption of motion of the AUV.}. Hence the cost caused by AUV's movement is given by
\begin{eqnarray}
\label{61}
CF=\sum\limits_{j=1}^{M}\chi{E_j},
\end{eqnarray}
Therefore, the profit of the MTUC can be obtained by
\begin{eqnarray}\small
\label{62}
\begin{aligned}
Pr&=Re-CT-CF\\
&=\sum\limits_{k=1}^{K}{\sum\limits_{i=1}^{{{N}_{k}}}({{\omega}_{{{k}_{i}}}T_{{{k}_{i}}}^{\rm S}+{{\lambda}_{{{k}_{i}}}}E_{{{k}_{i}}}^{\rm S}-CT^{\rm M}_{{k}_{i}}-CT^{\rm AS}_{{k}_{i}}})}-\sum\limits_{j=1}^{M}\chi{E_j}.
\end{aligned}
\end{eqnarray}

For maximizing the profit defined by Eq. (\ref{62}), we jointly optimize the computation offloading strategy $\boldsymbol{O}$, caching strategy $\boldsymbol{H}$, bandwidth allocation $\boldsymbol{R}$, computing resource allocation $\boldsymbol{F}$ and trajectory design strategy $\boldsymbol{Y}$. This optimization problem is formulated as
\begin{subequations}\label{OptimizationProblem}
\begin{eqnarray}
\label{63}
\begin{aligned}
\mathcal{P}1: \underset{\boldsymbol{O},\boldsymbol{H},\boldsymbol{R},\boldsymbol{F},\boldsymbol{Y}}{\mathop{\max}}\,
Pr\\
\end{aligned}
\end{eqnarray}
\begin{equation}\label{p1c1}
{s.t.}\quad\quad{{r}_{{{k}_{i}}}}=0, \: if \: {{o}_{{{k}_{i}}}}=0, \forall k\in \boldsymbol{K},i\in \boldsymbol{N}_k,
 \end{equation}
 \begin{equation}\label{p1c2}
{{r}_{{{k}_{i}}}}=0, \: if \: {{h}_{{{k}_{i}}}}=1,\forall k\in \boldsymbol{K},i\in \boldsymbol{N}_k,
\end{equation}
\begin{equation}\label{p1c3}
\sum\limits_{i=1}^{{{N}_{k}}}{{o}_{{{k}_{i}}}}({1-{h}_{{{k}_{i}}}}){{{r}_{{{k}_{i}}}}\le 1}, \forall k\in \boldsymbol{K},
\end{equation}
\begin{equation}\label{p1c4}
{{f}_{{{k}_{i}}}^m}=0, \: if \: {{o}_{{{k}_{i}}}}=0, \forall k\in \boldsymbol{K},i\in \boldsymbol{N}_k,
\end{equation}
\begin{equation}\label{p1c10}
\sum\limits_{j=1}^{M}{\sum\limits_{\xi=1}^{{{S}_{j}}}{{{Y}_{{{j}_{k}}}}[\xi]}}=1, \forall j \in \boldsymbol{M},  \forall k \in \boldsymbol{K},
\end{equation}
\begin{equation}\label{p1c5}
\sum\limits_{i=1}^{{{N}_{k}}}{{o}_{{{k}_{i}}}}{{{f}_{{{k}_{i}}}^m}\le 1}, \forall k\in \boldsymbol{K},
\end{equation}
\begin{equation}\label{p1c8}
\sum\limits_{k=1}^{K}{\sum\limits_{i=1}^{{{N}_{k}}}{{h_{{k}_{i}}}{{Z}_{{{k}_{i}}}}}}\le {{C}_{e}},
\end{equation}
\begin{equation}\label{p1c9}
\boldsymbol{P}_{j}^{\textrm{A}}[S_j+1]=\boldsymbol{P}_{j}^{\textrm{A}}[0], \forall j \in \boldsymbol{M},
\end{equation}
\begin{equation}\label{p1c11}
\sum\limits_{\xi=1}^{{{S}_{j}}}{\sum\limits_{k=1}^{K}{{{Y}_{{{j}_{k}}}}}}[\xi]={{S}_{j}}, \forall j \in \boldsymbol{M},
\end{equation}
\begin{equation}\label{p1c12}
\sum\limits_{j=1}^{M}{{{S}_{j}}}=K, \forall j \in \boldsymbol{M},
\end{equation}
\begin{equation}\label{p1c13}
{{T}^{\textrm{AT}}_{\max }}-{{T}^{\textrm{AT}}_{\min }}\le \varepsilon,
\end{equation}
\begin{equation}\label{p1c7}
{{h}_{{{k}_{i}}}}\le {{o}_{{{k}_{i}}}}.
\end{equation}
\end{subequations}

As for $\mathcal{P}1$, we have following proposition:
\begin{proposition}\label{proposition1}
$\mathcal{P}1$ is NP-hard, and we cannot find an optimal solution in polynomial time.
\end{proposition}

\begin{proof}
The detail proof is provided in Appendix. \ref{ProofProposition1}.
\end{proof}

\section{Deep Reinforcement Learning Solution}\label{CC}
Since the problem formulated is non-convex and NP-hard, which is generally intractable for conventional optimization methods, therefore, we introduce A3C \cite{9385791}, an efficient distributed deep reinforcement learning approach,  for solving $\mathcal{P}1$. Briefly, A3C combines the advantages of both value-based and policy-based reinforcement learning algorithms, which can deal with both continuous and discrete valued problems and implement an asynchronous update for improving learning efficiency.
\subsection{Modeling of Deep Reinforcement Learning Environment}\label{A}
Specifically, we need to transform $\mathcal{P}1$ to an MDP firstly, which consists of state space, action space, policy, state transition matrix function, and reward function.

{\bfseries State Space}: At each episode $\vartheta$, the state $s(\vartheta) \in \mathcal{S}$ includes the following parts:
\begin{itemize}
	\item The coordinates of AUVs at episode $\vartheta$:\\ $\left\{\boldsymbol{P}_{j}^{\textrm{A}}(\xi,\vartheta), j \in \boldsymbol{M}, \xi \in \boldsymbol{S_{j}}\right\}$;
	\item The offloading strategy at episode $\vartheta-1$:\\$\left\{o_{k_i}(\vartheta-1), k \in \boldsymbol{K}, i \in \boldsymbol{N}_k\right\}$;
	\item The caching strategy at episode $\vartheta-1$:\\$\left\{h_{k_i}(\vartheta-1), k \in \boldsymbol{K}, i \in \boldsymbol{N}_k\right\}$ ;
	\item The bandwidth allocation at episode $\vartheta-1$:\\$\left\{r_{k_i}(\vartheta-1), k \in \boldsymbol{K}, i \in \boldsymbol{N}_k\right\}$;
	\item The computing resource allocation at episode $\vartheta-1$:\\$\left\{f_{k_i}^{m}(\vartheta-1), k \in \boldsymbol{K}, i \in \boldsymbol{N}_k\right\}$;
	\item The trajectory design strategy at episode $\vartheta-1$:\\$\left\{Y_{j_k}(\xi, \vartheta-1), j \in \boldsymbol{M}, \xi \in \boldsymbol{S_{j}}, k \in \boldsymbol{K} \right\};$
\end{itemize}
Hence, the state at episode $\vartheta$ can be summarized as
\begin{equation}
\label{64}
\begin{aligned}
& s(\vartheta)=\left\{\boldsymbol{P}_{j}^{\textrm{A}}(\xi,\vartheta), o_{k_i}(\vartheta-1), h_{k_i}(\vartheta-1), r_{k_i}(\vartheta-1), \right. \\ &\left. f_{k_i}^{m}(\vartheta-1), Y_{j_k}(\xi, \vartheta-1), j \in \boldsymbol{M}, \xi \in \boldsymbol{S_{j}}, k \in \boldsymbol{K}, i \in \boldsymbol{N}_k\right\}.
\end{aligned}
\end{equation}

{\bfseries Action Space}: At each episode $\vartheta$, the agent selects an action $a(\vartheta) \in \mathcal{A}$ according
to the observed state  $s(\vartheta)$, where $a(\vartheta)$ consists of the following parts:
\begin{itemize}
	\item The offloading strategy at episode $\vartheta$:\\$\left\{o_{k_i}(\vartheta), k \in \boldsymbol{K}, i \in \boldsymbol{N}_k\right\}$;
	\item The caching strategy of task at episode $\vartheta$:\\$\left\{h_{k_i}(\vartheta), k \in \boldsymbol{K}, i \in \boldsymbol{N}_k\right\}$ ;
	\item The bandwidth allocation at episode $\vartheta$:\\$\left\{r_{k_i}(\vartheta), k \in \boldsymbol{K}, i \in \boldsymbol{N}_k\right\}$;
	\item The computing resource allocation at episode $\vartheta$:\\$\left\{f_{k_i}^{m}(\vartheta), k \in \boldsymbol{K}, i \in \boldsymbol{N}_k \right\}$;
	\item The trajectory design strategy at episode $\vartheta$:\\$\left\{Y_{j_k}(\xi, \vartheta), j \in \boldsymbol{M}, \xi \in \boldsymbol{S_{j}}, k \in \boldsymbol{K} \right\} ;$
\end{itemize}
Hence, the action at episode $\vartheta$ can be formulated as
\begin{eqnarray}\small
\label{64}
\begin{aligned}
& a(\vartheta)=\left\{o_{k_i}(\vartheta), h_{k_i}(\vartheta), r_{k_i}(\vartheta),f_{k_i}^{m}(\vartheta), Y_{j_k}(\xi, \vartheta)\right. \\ &\left. f_{k_i}^{m}(\vartheta-1), Y_{j_ k}(\xi, \vartheta-1), j \in \boldsymbol{M}, \xi \in \boldsymbol{S_{j}}, k \in \boldsymbol{K}, i \in \boldsymbol{N}_k\right\}.
\end{aligned}
\end{eqnarray}

{\bfseries Policy}: Let $\pi(a \mid s )=\mathcal{P}\left(a \mid s\right)$ denote the policy function, which is a probability distribution based on the observed state to make a decision to select an action.

{\bfseries State Transition Function}: Let $\mathcal{P}\left[s(\vartheta+1) \mid s(\vartheta), a(\vartheta)\right]$ be the transition probability at each episode, which is the probability of entering into the state $s(\vartheta+1)$ after executing action $a(\vartheta)$ at the observed state $s(\vartheta)$.

{\bfseries Reward Function}: The reward function is the objective of  Eq. $\eqref{63}$ for the sake of maximizing the profit of the MTUC framework, which is represented as
\begin{eqnarray}
\begin{aligned}
 r&\left(s(\vartheta), a(\vartheta)\right) = \\ &\sum\limits_{k=1}^{K}{\sum\limits_{i=1}^{{{N}_{k}}}({{\omega}_{{{k}_{i}}}T_{{{k}_{i}}}^{\rm S}+{{\lambda}_{{{k}_{i}}}}E_{{{k}_{i}}}^{\rm S}-CT^{\rm M}_{{k}_{i}}-CT^{\rm AS}_{{k}_{i}}})} -\sum\limits_{j=1}^{M}\chi{E_j}.
\end{aligned}
\end{eqnarray}
\subsection{A3C-Based Joint Optimization Algorithm}\label{B}
 \begin{figure*}
  \centering
  \includegraphics[width=15cm]{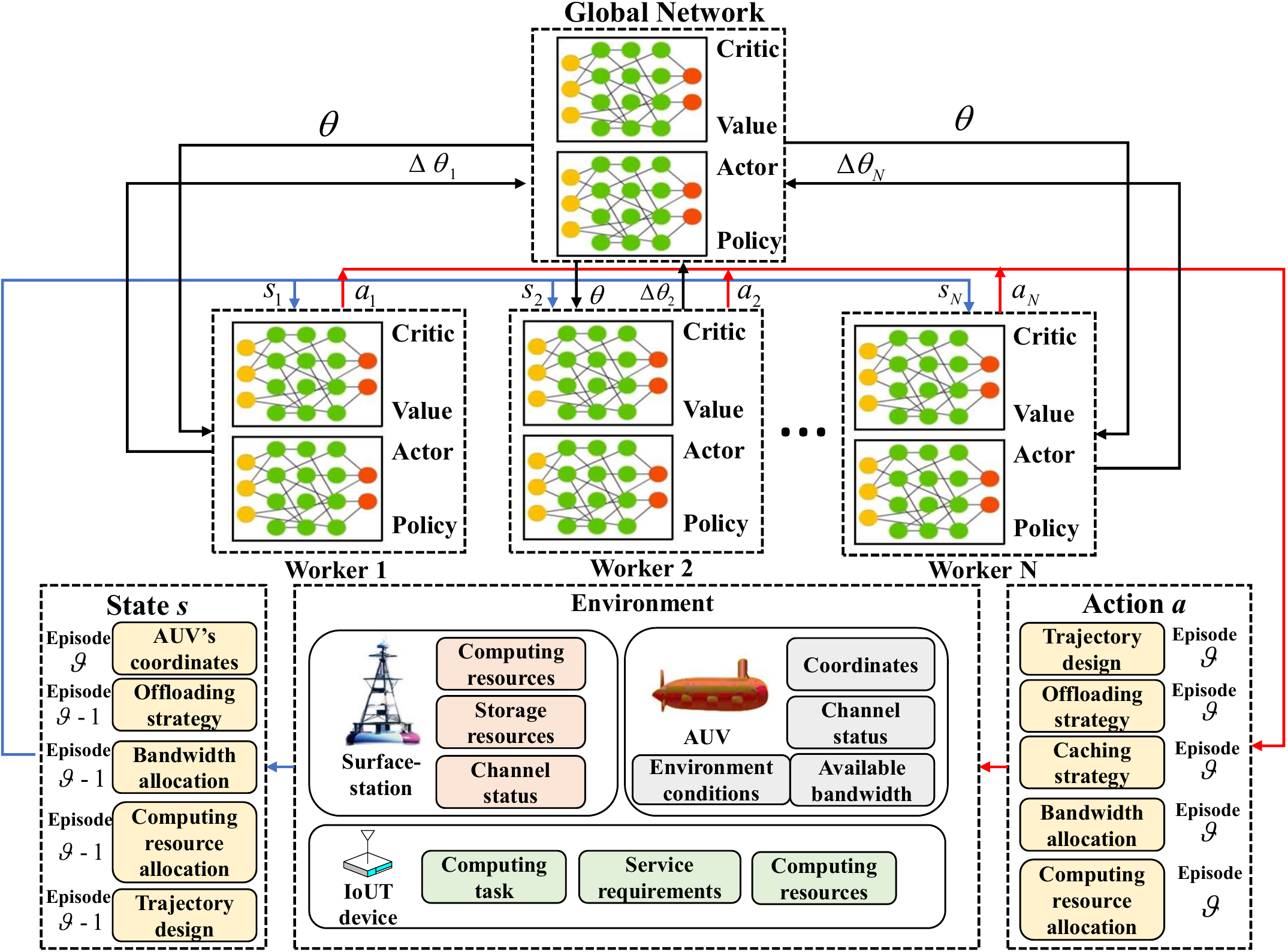}
  \caption{The architecture of A3C-based joint optimization algorithm.}\label{A3C_architecture}
\end{figure*}
Here, A3C is adopted to deal with the large-scale optimization problem  formulated. The architecture of the A3C-based joint optimization algorithm is shown in Fig. \ref{A3C_architecture}. In contrast to the traditional deep reinforcement learning method, A3C can realize efficient distributed asynchronous learning.  In the A3C-based joint optimization algorithm, the agent consists of a global network and multiple workers. Both the global network and the workers have the same network architecture, which is composed of two neural networks, namely the policy network (actor) with parameter ${\theta}_A$ and the value network (critic) with parameter ${\theta}_C$. The workers learn in parallel by interacting with their environments separately to compute their new gradients and send them to the global networks, when reaching the terminal state or the maximum number of iterations. Instead of interacting with the environment directly, the global network is only responsible for updating the global network parameters with the gradient fetched from the workers and distributing the global network parameters to each worker at regular intervals.

Specifically, in each episode, the estimated state value predicted by the value network is denoted by $V\left[s(\vartheta) ; \theta_{C}\right]$. The agent executes an action $a(\vartheta)$ according to the policy $\pi\left[a(\vartheta) \mid s(\vartheta)\right]$ at the current state $s(\vartheta)$, and then the environment will change to the next state $s(\vartheta+1)$ and generate a reward $r(\vartheta)$. The state value function of A3C is represented as \cite{wang2020on}
\begin{eqnarray}
\label{68}
\begin{aligned}
V\left(s(\vartheta) ; \theta_{C}\right)=E\left[\sum_{c=0}^{\infty} \Psi^{c} r(\vartheta+c) \right],
\end{aligned}
\end{eqnarray}
where $\Psi$ is the discount factor, which denotes how future rewards affect the current state value. A3C employs a $\mathbb{K}$-step reward for updating the parameters, which can be represented as
\begin{eqnarray}
\label{68}
\begin{aligned}
R(\vartheta)=\sum_{l=0}^{\mathbb{K}-1} \Psi^{l} r(\vartheta+l)+\Psi^{\mathbb{K}} V\left(s_{\vartheta+\mathbb{K}} ; \theta_{C}\right),
\end{aligned}
\end{eqnarray}
where $\mathbb{K}$ is the number of time steps required for calculating $\mathbb{K}$-step returns. Aiming for reducing the estimation variance and improving the decision-making capability of the agent, we define the advantage function $\hat A(\vartheta)$ as follows:
\begin{eqnarray}
\label{68}
\begin{aligned}
\hat A\left[s(\vartheta), a(\vartheta) ; {\theta}_A, \theta_{C}\right]=R(\vartheta)-V\left[s(\vartheta) ; \theta_{C}\right].
\end{aligned}
\end{eqnarray}
Furthermore, the loss function of the actor is represented as
\begin{equation}
\label{68}
\begin{aligned}
J_{\pi}({\theta}_A)= \log \pi\left[a(\vartheta) \mid s(\vartheta) ; {\theta}_A\right]\hat A(\vartheta)+\Theta H\left(\pi\left[s(\vartheta) ; {\theta}_A\right]\right),
\end{aligned}
\end{equation}
where $H\left(\pi\left[s(\vartheta) ; {\theta}_A\right]\right)$ is an entropy item introduced for encouraging exploration and for avoiding to fall into a local optimum,  while $\Theta$ manages the strength of the entropy regularization. By contrast, the loss function of the critic network is denoted by
\begin{eqnarray}
\label{68}
\begin{aligned}
J_{C}({\theta}_A)={\hat A(\vartheta)}^{2}.
\end{aligned}
\end{eqnarray}
As the updating process, the accumulated gradient of the policy network is calculated as
\begin{eqnarray}
\label{DA}
\begin{aligned}
d {\theta}_A \leftarrow d {\theta}_A &+\nabla_{{\theta}_A} \log \pi\left[a(\vartheta) \mid s(\vartheta) ; {\theta}_A\right]\hat A(\vartheta) \\
&+\delta \nabla_{{\theta}_A} H\left(\pi\left[s(\vartheta) ; {\theta}_A\right]\right),
\end{aligned}
\end{eqnarray}
while the accumulated gradient of the value network is calculated as
\begin{eqnarray}
\label{DC}
\begin{aligned}
d \theta_{C} \leftarrow d \theta_{C}+\frac{\partial{\hat A(\vartheta)}^{2}}{\partial \theta_{C}}.
\end{aligned}
\end{eqnarray}

To train the A3C framework effectively, the RMSProp algorithm \cite{wang2020on} is adopted, which can significantly improve the speed of gradient descent. The estimated gradient relying on the RMSProp algorithm can be formulated as
\begin{eqnarray}
\label{68}
\begin{aligned}
\Upsilon =\Lambda \Upsilon+(1-\Lambda) (\Delta {\theta})^{2},
\end{aligned}
\end{eqnarray}
where $\Delta {\theta}$ represents the accumulated gradients of the loss function of the policy or value networks,  while $\Lambda$ is the momentum.
Relying on Eq. (\ref{68}), we update the parameters of the policy and value networks by
\begin{eqnarray}
\label{UpA}
\begin{aligned}
{\theta}_A \leftarrow {\theta}_A-\Xi \frac{\Delta {\theta}_A}{\sqrt{\Upsilon+\epsilon}}
\end{aligned}
\end{eqnarray}
and
\begin{eqnarray}
\label{UpC}
\begin{aligned}
{\theta}_C \leftarrow {\theta}_C-\Xi \frac{\Delta {\theta}_C}{\sqrt{\Upsilon+\epsilon}},
\end{aligned}
\end{eqnarray}
respectively, where $\epsilon$ is a tiny positive step, while $\Xi$ is the learning rate. The procedure designed is summarized in Algorithm 1.
\begin{algorithm}[t]
\caption{Asynchronous advantage actor-critic Algorithm}
\label{}
\begin{algorithmic}
\small
\STATE Initialize the maximum counters $\mathcal{T}_{max}$, $\vartheta_{max}$, and all the parameters as shown in Table \ref{ParaSettings}, respectively.
\STATE Initialize the global policy network and global value network with parameters $ \theta_{A}$ and $ \theta_{C}$.
\STATE Initialize global shared counter as $\mathcal{T}= 0 $ and thread-specific counter as $\vartheta = 1$.
\STATE Initialize the thread-specific policy network parameters ${\theta}_A^{\prime}$ and value network parameters $\theta_{C}^{\prime}$.
\FOR {$\mathcal{T} < \mathcal{T}_{max}$}
\FOR {each worker}
\STATE Initialize the gradients of  agent as $d \theta_{A} = 0$ and $d \theta_{C}=0$.
\STATE Synchronous parameters of each worker with global parameters ${\theta}_A^{\prime}={\theta}_A$ and ${\theta}_C^{\prime} = {\theta}_C$.

\FOR {$\vartheta \le \vartheta_{max}$}
\STATE Obtain the state $s(\vartheta)$.
\STATE Perform $a(\vartheta)$ relying on the policy $\pi(a(\vartheta) \mid s(\vartheta) ; {\theta}_A^{\prime})$.
\STATE Obtain reward $r(\vartheta)$ and new state $s(\vartheta+1)$.
\STATE $\vartheta = \vartheta + 1$.
\ENDFOR
\STATE $\hat V=\left\{\begin{array}{ll}0, & \text { for terminal state } \\ V\left(s(\vartheta), \theta_{v}^{\prime}\right), & \text{ for non-terminal state }\end{array}\right.$
\FOR {$\vartheta = \vartheta_{max}$}
\STATE $\hat V = r(\vartheta) + \Psi \hat V$
\STATE Obtain the accumulate gradient with respect to $ \theta_{A}'$ by Eq. (\ref{DA});
\STATE Obtain the accumulate gradient with respect to $ \theta_{C}'$ by Eq. (\ref{DC});
\ENDFOR
\STATE Update $ \theta_{A}$ and $ \theta_{C}$ according to Eq. (\ref{UpA}) and Eq. (\ref{UpC}).
\STATE $\mathcal{T} = \mathcal{T}+1$
\ENDFOR
\ENDFOR
\normalsize
\end{algorithmic}
\end{algorithm}
\section{Simulation Results}
In this section, we provide the experimental results for validating the superiority of our proposed scheme. Unless specified, otherwise, the number of the AUVs is set to 4, while the numbers of the DGs and IoUT devices are set to 15 and 190, respectively. The main parameters are summarized in Table \ref{ParaSettings}.

\begin{table}[t]
\centering
\caption{Values of main parameters}
\label{ParaSettings}
\begin{tabular}{|c|c|c|c|}
\hline
\textbf{Parameter} & \textbf{Value}     & \textbf{Parameter} &  \textbf{Value}  \\ \hline \hline
$f$                                 & 30 kHz                              & ${P^{\rm D}_{\rm tr}}$              & 30 mW                           \\ \hline
$s$                                 & 0.5                                 & ${P^{\rm A}_{\rm tr}}$              & 36 mW                           \\ \hline
$w$                                 & 0                                   & $B_{\rm H}$                         & 10 kHz                           \\ \hline
$H$                                 & 200 m                               & $B_{\rm L}$                         & 10 kHz                           \\ \hline
$h_0$                               & 10 m                                & $\mathcal{V}$                       & 1                               \\ \hline
$d_0$                               & 20 m                                & $V_k$                               & 5 knot                          \\ \hline
$\Omega_0$                          & 8                                   & $C_a$                               & 0.0314                          \\ \hline
$\rho_L$                            & 1020 $ \rm{ kg/m ^{3}}$                    & $\alpha$                            & 100                             \\ \hline
$Z_{ki}$                            & $\mathcal{U}[10^{5},3*10^{5}]$ bit  & $\beta$                             & 100                             \\ \hline
$f_{ki}$                            & $\mathcal{U}[1,4]$ GHz              & $\phi$                              & 100                             \\ \hline
$\alpha_{ki}$                       & $\mathcal{U}[1500,2000]$ cycles/bit & $\varepsilon$                       & 2 s                             \\ \hline
$C_e$                               & 100 Mb                              & $r_0$                               & 100 m                           \\ \hline
$C_d$                               & 0.117                               & $\sigma$                            & 3                               \\ \hline
$k_s$                               & 1.5                                 & $\mu$                               & $1.25 \times 10 ^{-26}$         \\ \hline
$\Gamma_{s}$                        & 1                                   & $\zeta$                             & 0.8                             \\ \hline
$\Gamma_{b}$                        & 0.0139                              & $\eta$                              & 0.2                             \\ \hline
$H_1$                               & 180 m                               & $H_2$                               & 190 m                           \\ \hline
$\omega_{k_i} $                     & $\mathcal{U}[10,20]$                & $\lambda_{k_i}$                     & $\mathcal{U}[1,2]$              \\ \hline
$\varrho $                          & 1                                 & $\chi $                             & 2                               \\ \hline
\end{tabular}
\end{table}
\subsection{Impact of the Hostile Underwater Environment on the System}
\begin{figure*}[t]\centering
\subfigure[Providing service by a single AUV without environmental awareness.]{\includegraphics[height=3.9cm, width=4.2cm]{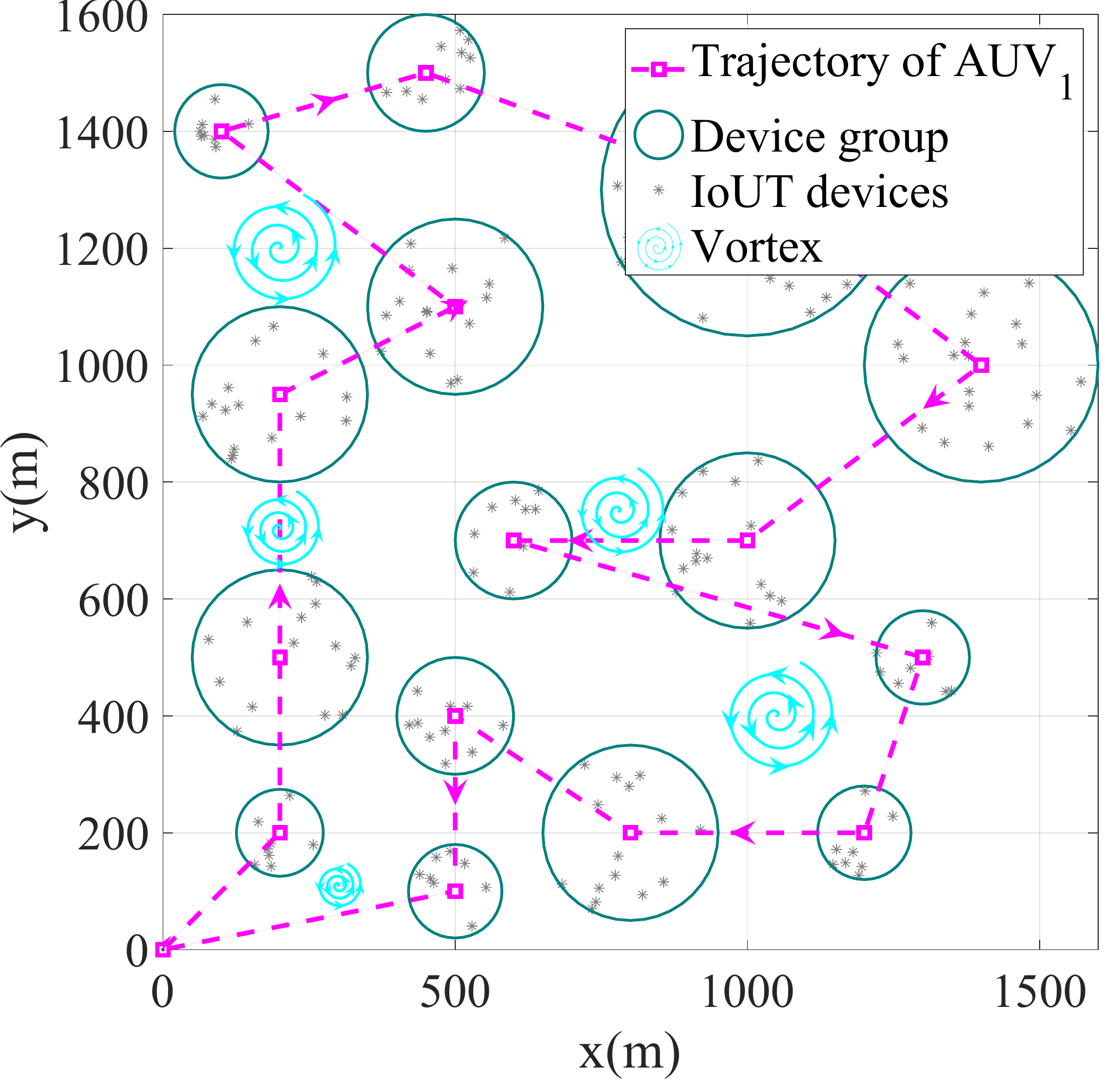}  \label{Trajectory1}}
\subfigure[Providing service by 2 AUVs without environmental awareness.]{  \includegraphics[height=3.9cm, width=4.2cm]{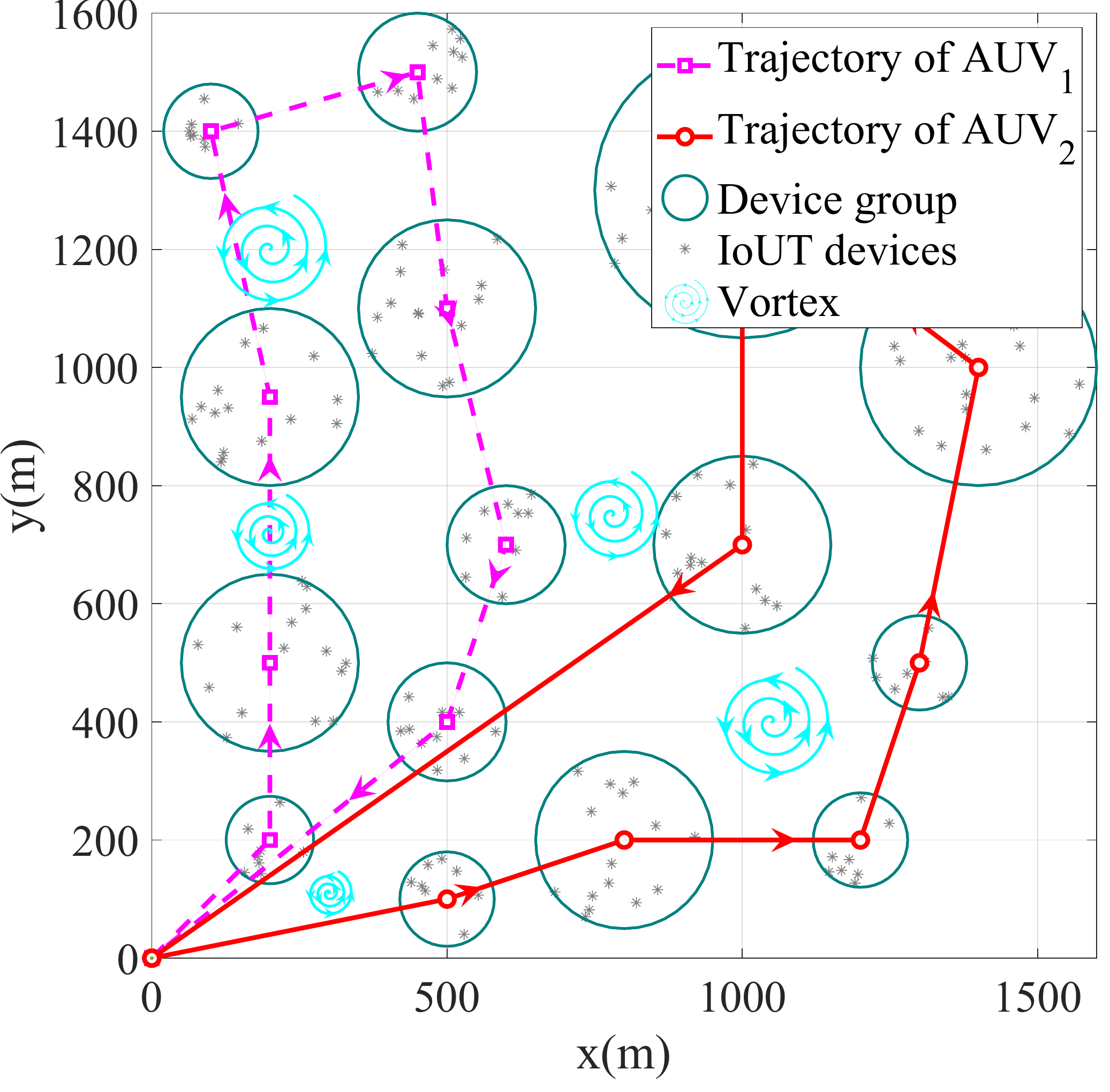} \label{Trajectory2} }
\subfigure[Providing service by 3 AUVs without environmental awareness.]{  \includegraphics[height=3.9cm, width=4.2cm]{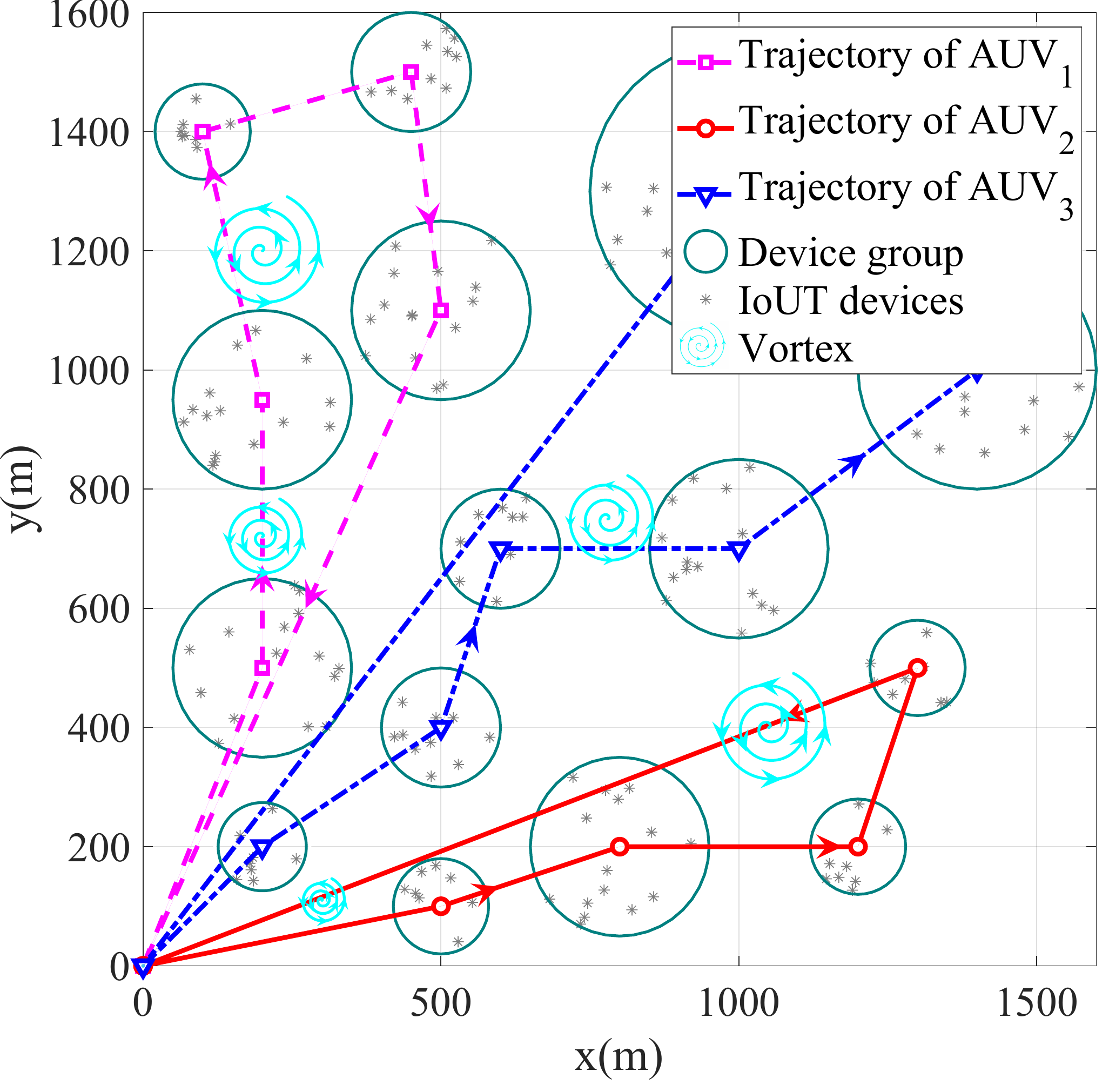} \label{Trajectory3} }
\subfigure[Providing service by  4 AUVs without environmental awareness.]{  \includegraphics[height=3.9cm, width=4.2cm]{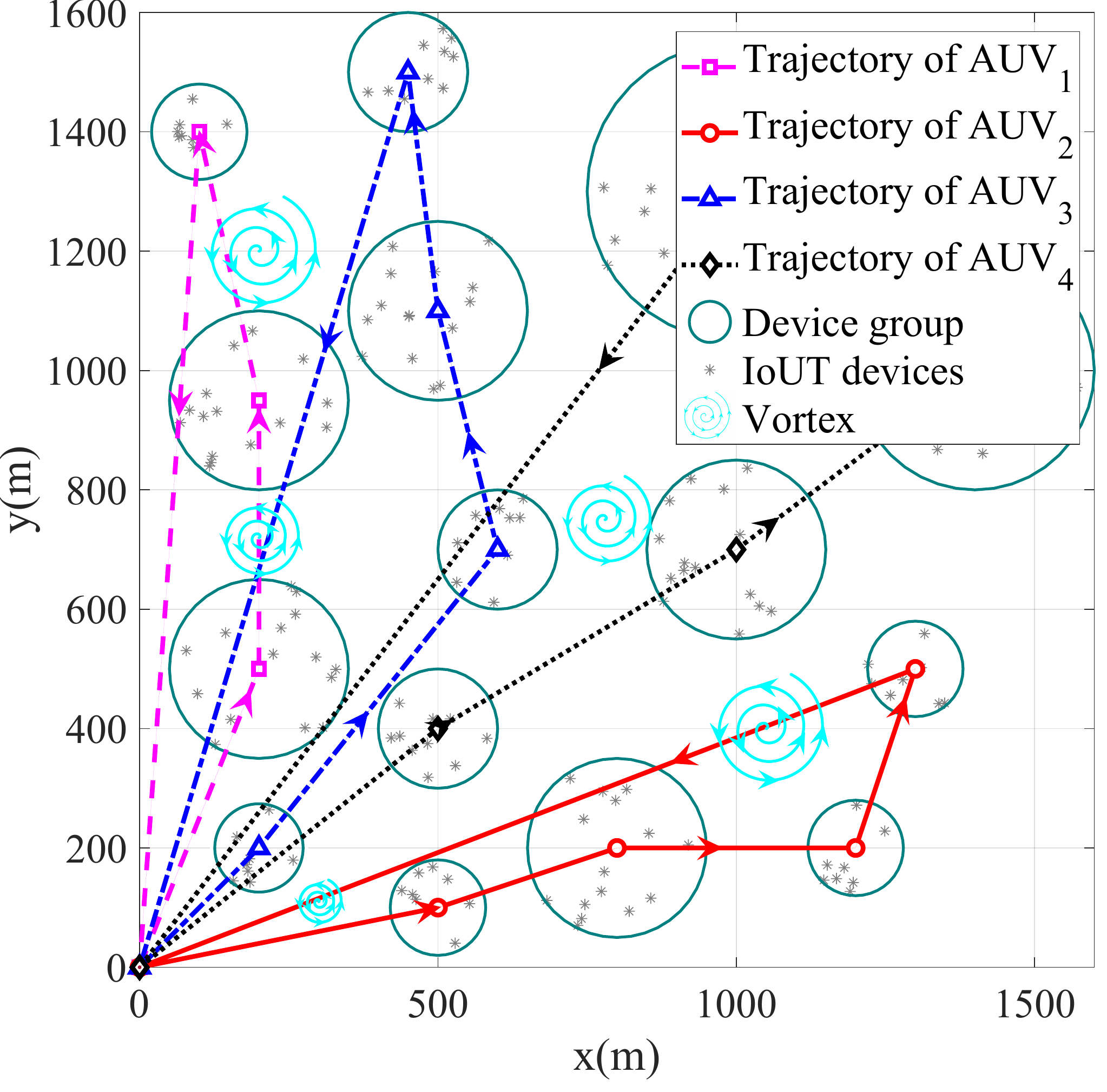}\label{Trajectory4}  }
\subfigure[Providing service by a single AUV with environmental awareness.]{  \includegraphics[height=3.9cm, width=4.2cm]{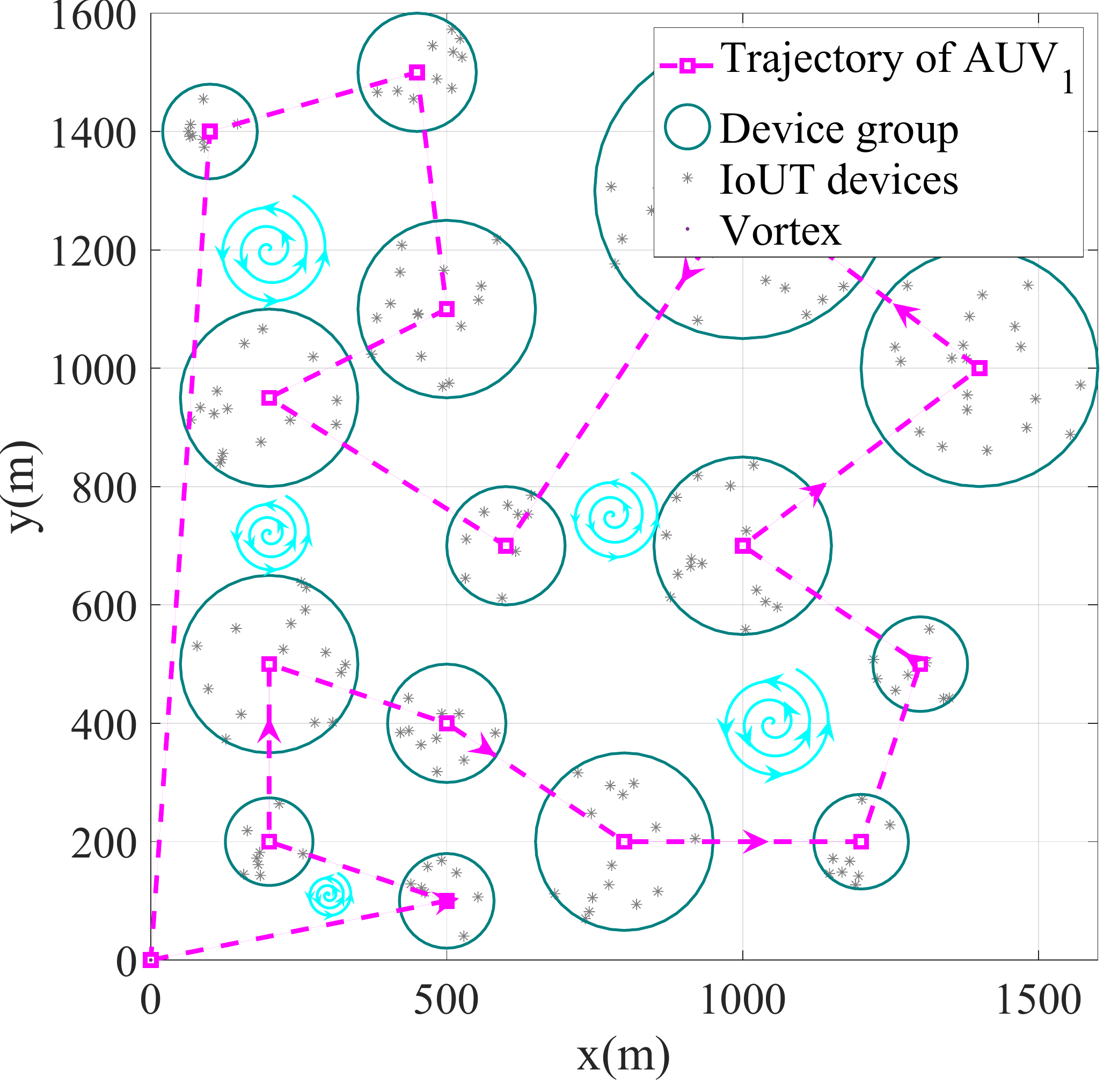} \label{Trajectory5} }
\subfigure[Providing service by  2 AUVs with environmental awareness.]{  \includegraphics[height=3.9cm, width=4.2cm]{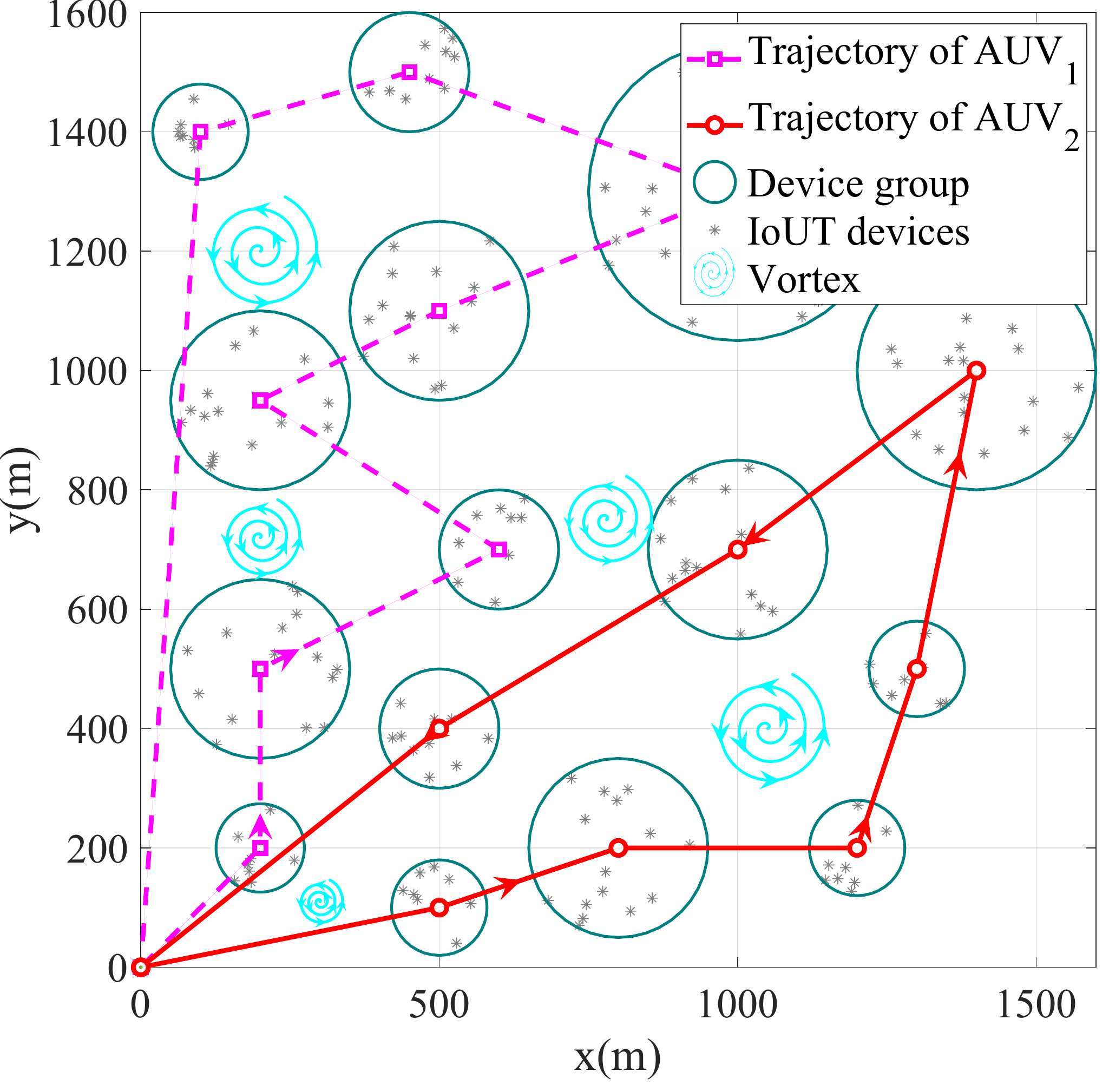} \label{Trajectory6} }
\subfigure[Providing service by  3 AUVs with environmental awareness.]{  \includegraphics[height=3.9cm, width=4.2cm]{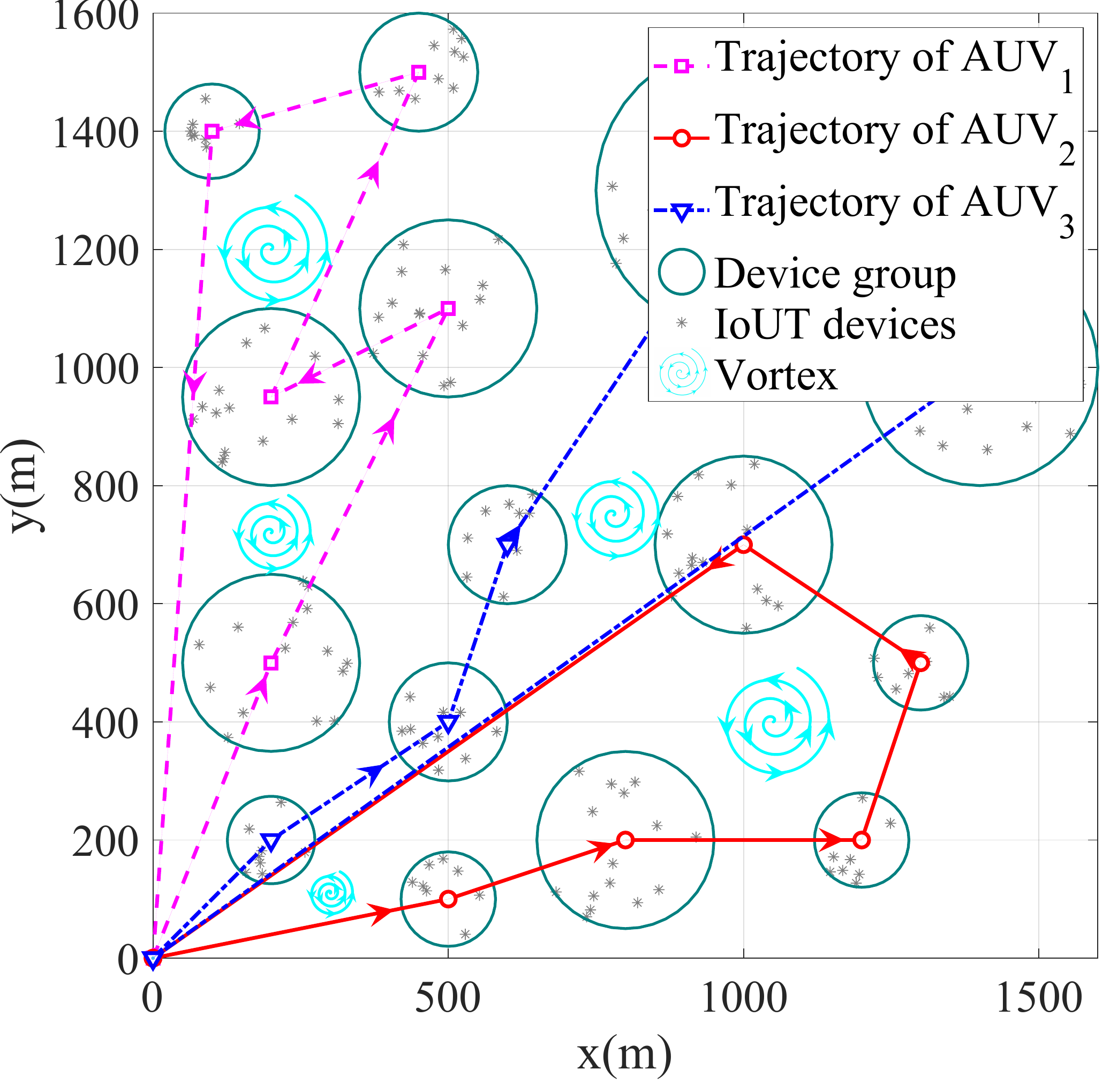} \label{Trajectory7} }
\subfigure[Providing service by  4 AUVs with environmental awareness.]{  \includegraphics[height=3.9cm, width=4.2cm]{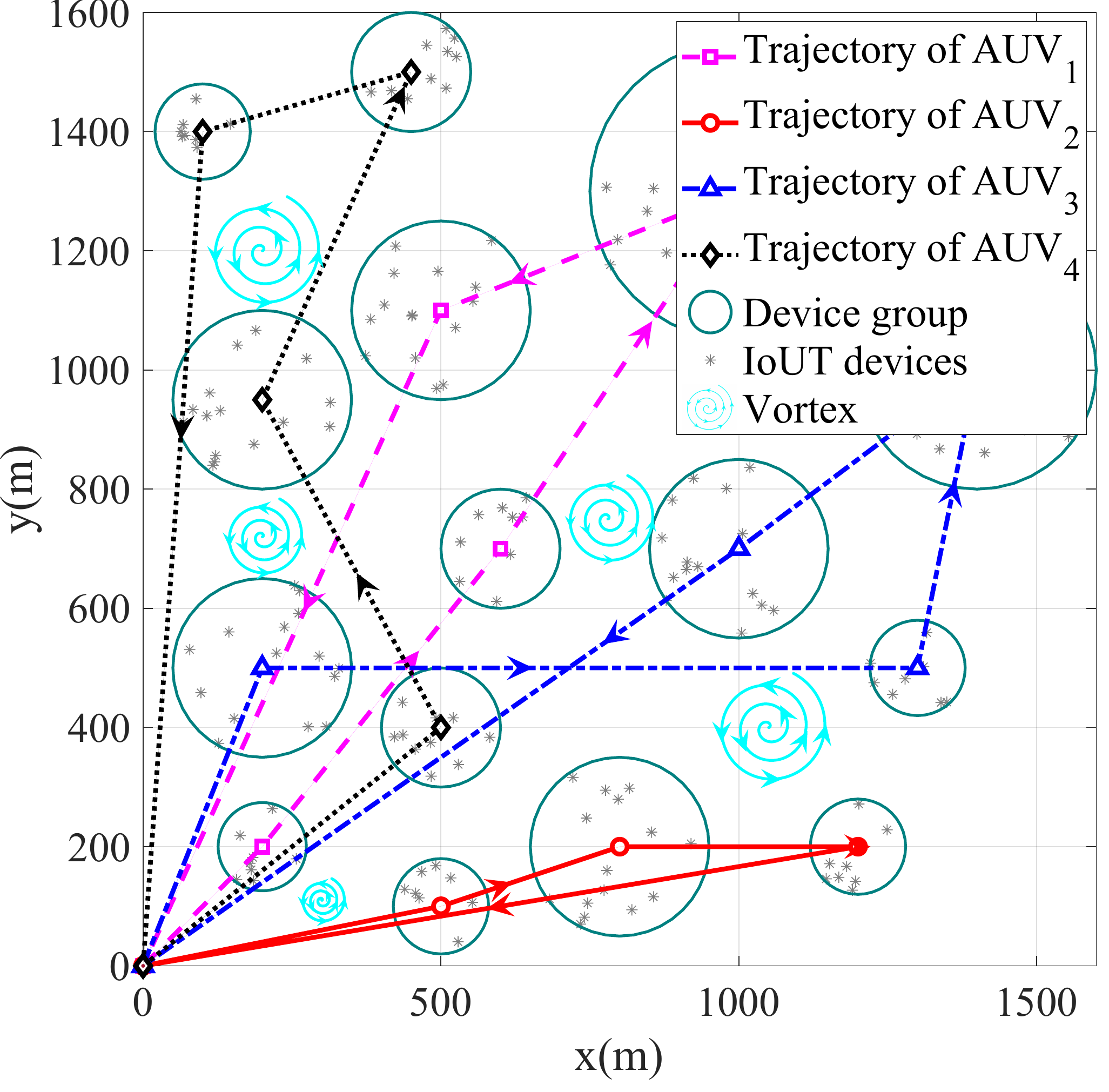}\label{Trajectory8}  }
 \caption{Comparison between environment-agnostic and environment-aware trajectory design.}\label{Trajectory}
\end{figure*}
Fig. \ref{Trajectory} shows the difference between the trajectory design with and without environmental awareness. Observe that each AUV starts from the origin, providing services to the DGs assigned, and then returns to the starting point for recharging after completing one cycle. Furthermore, as we can observe, compared to the AUVs in Fig. \ref{Trajectory1}-\ref{Trajectory4},  the AUVs in Fig. \ref{Trajectory5}-\ref{Trajectory8} relying on environmental awareness can select the optimal trajectories without vortex, which can avoid the extra energy consumption of the vortex and yield a high profit for the MTUC framework. Although sometimes the AUV relying on environmental awareness selects a longer path than that without environmental awareness, the profit of the whole system still settles on the global optimum.

\begin{figure}
  \centering
  \includegraphics[width=7cm]{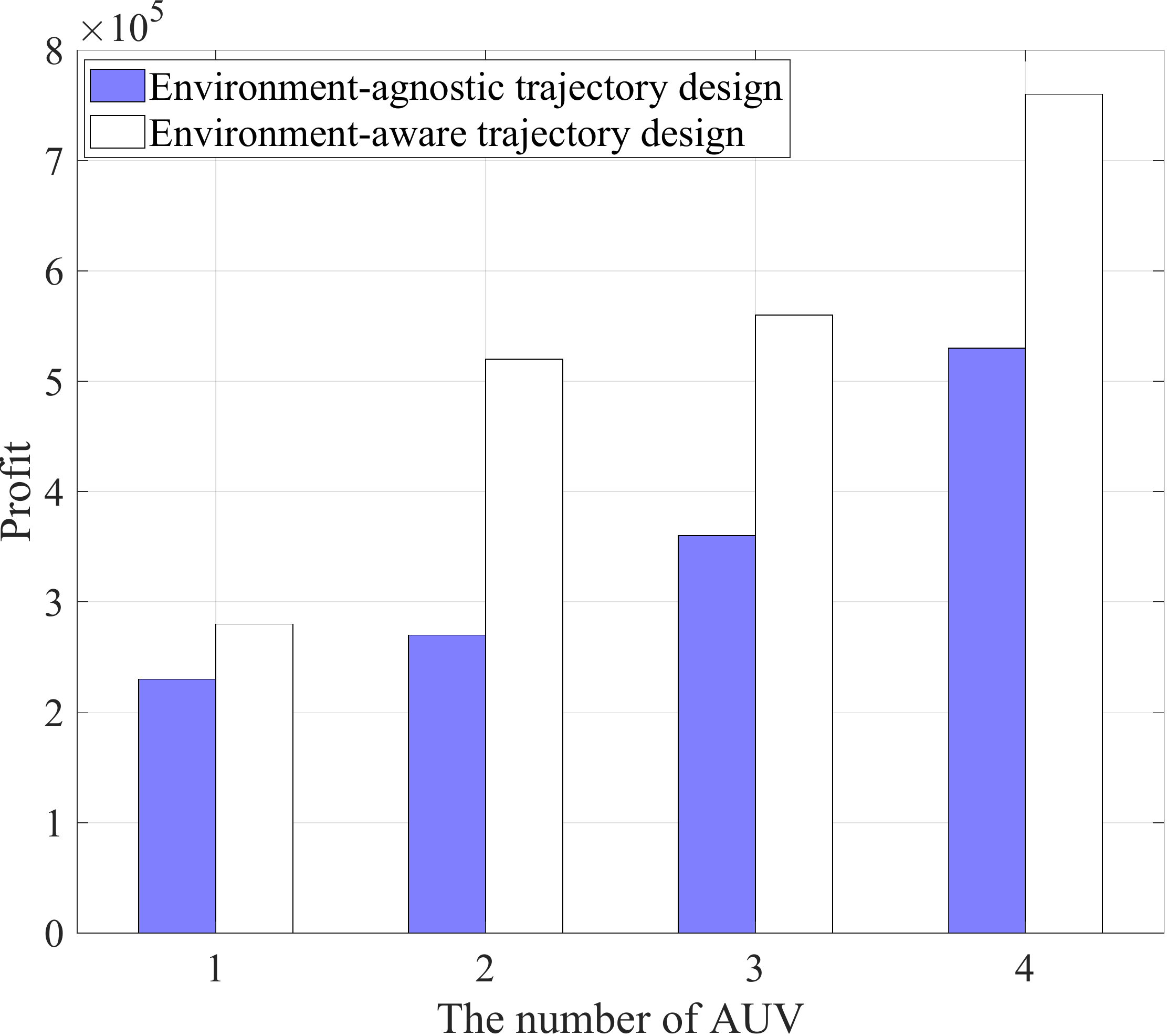}\\
  \caption{Comparison of the profit between environment-aware and environment-agnostic trajectory design versus the number of AUVs.} \label{PTraEnv}
\end{figure}
In Fig. \ref{PTraEnv}, we show the profit comparison between environment-aware and environment-agnostic trajectory design versus the number of AUVs, corresponding to the results shown in Fig. \ref{Trajectory}. Observe that the environment-aware trajectory design outperforms its agnostic counterpart. Furthermore, as the number of the AUVs increases, the profit of the whole system increases, because the collaboration of multiple AUVs exhibits more flexibility than a single AUV. However, we will conjecture that having a higher number of AUVs does not necessarily result in a higher profit for the system.
 \begin{figure}
  \centering
  \includegraphics[width=7cm]{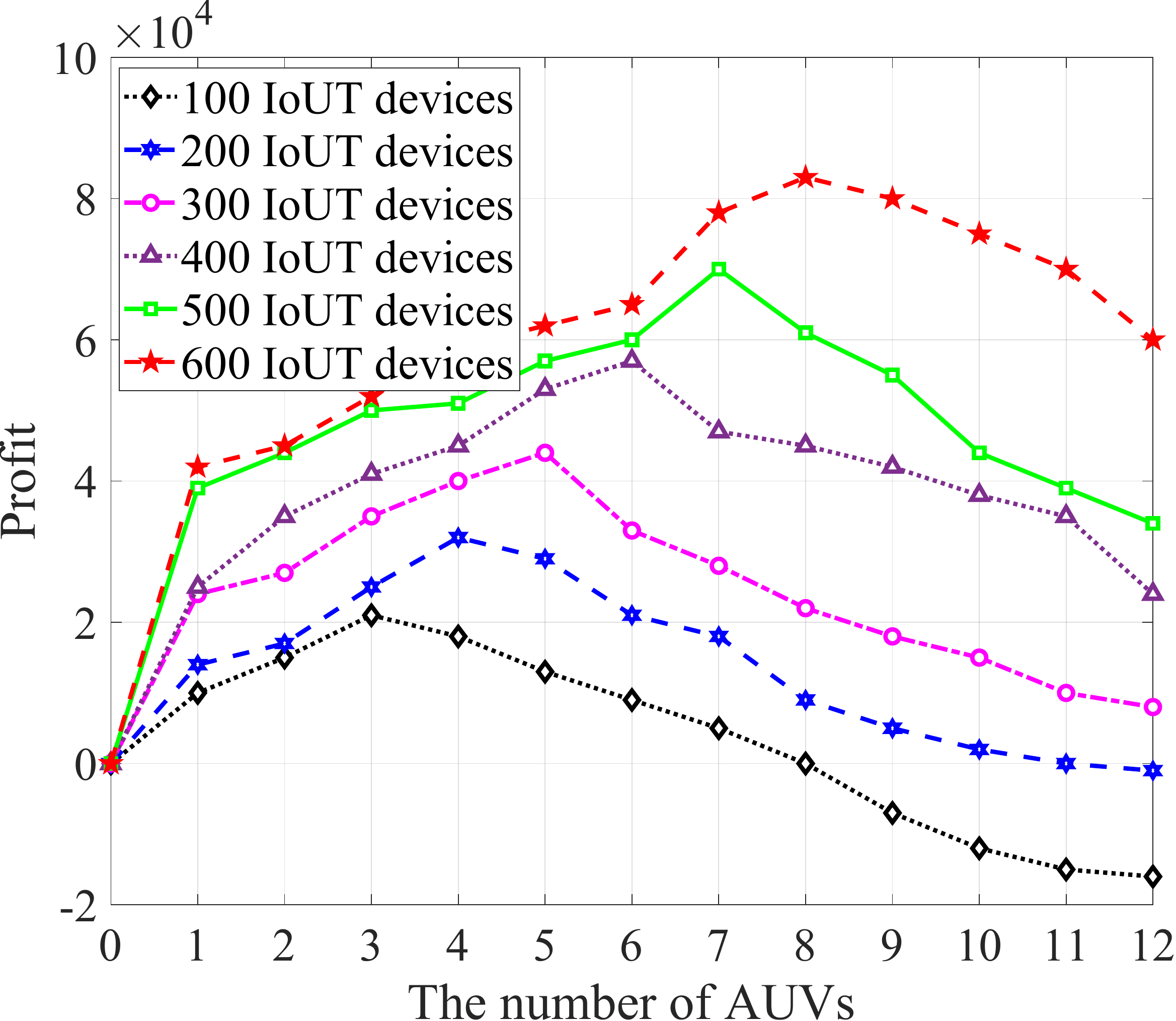}\\
  \caption{The profit versus the number of AUVs serving different numbers of IoUT devices.} \label{IoUT_AUV}
\end{figure}

Fig. \ref{IoUT_AUV} portrays the profit versus the number of AUVs serving different numbers of IoUT devices. Observe that there is always an optimal solution for the number of AUVs for serving a given number of IoUT devices. For example, for 300 IoUT devices, employing 5 AUVs to provide services achieves the highest profit. This phenomenon can provide us with a tangible philosophy for guiding the AUV deployment. Furthermore, we can observe in Fig. \ref{IoUT_AUV} that if all other conditions remain the same, then increasing the number of devices increases the benefit of the system. The reason for this is that when the number of devices increases, assigning the same energy consumption to the AUV's movement can support more IoUT devices, thereby obtaining higher revenue and further improving the profit.
\subsection{Impact of Different Resource Allocation Schemes on the System's Profit}

\begin{figure}
\centering
\includegraphics[width=7cm]{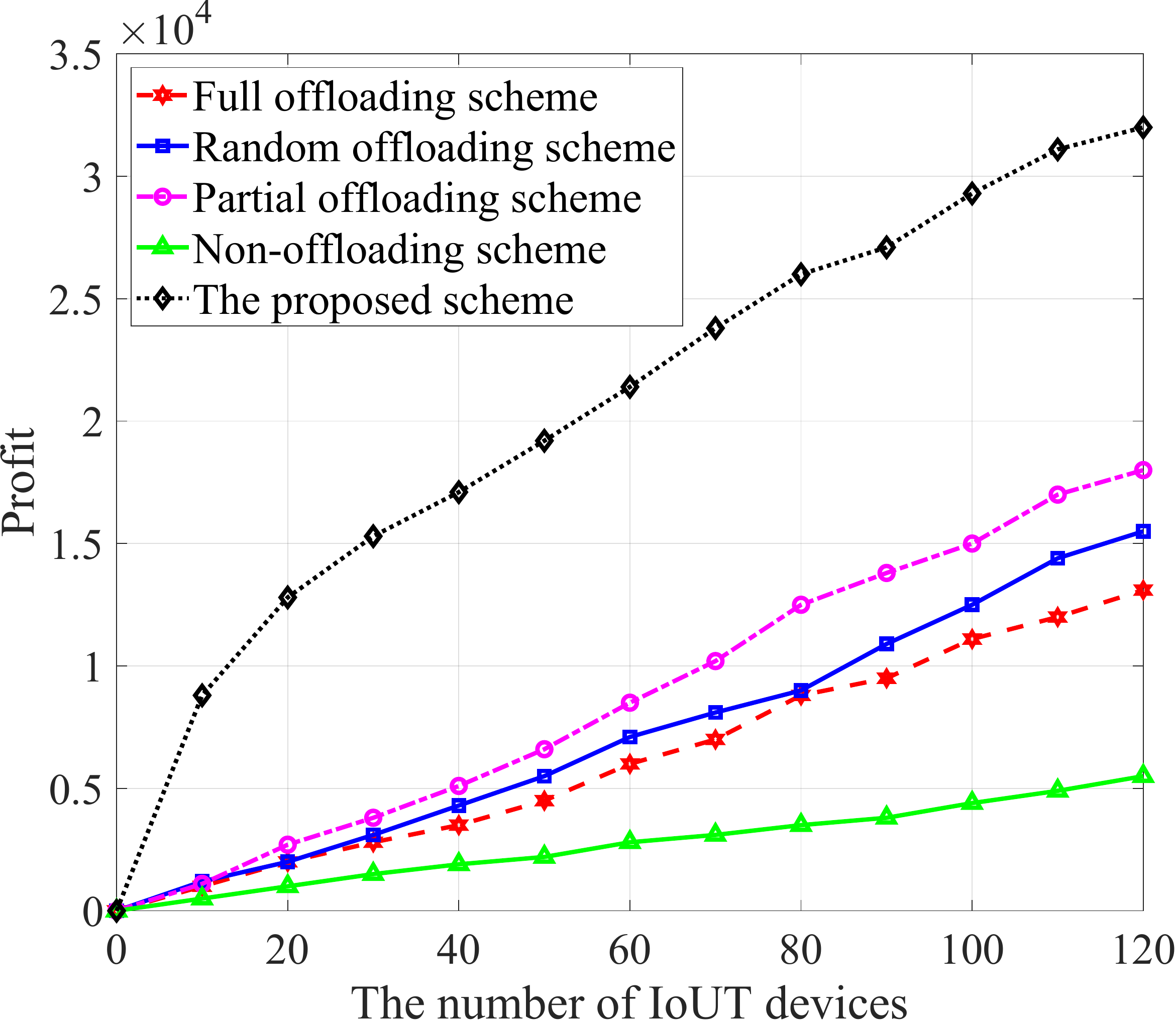}\\
\caption{The profit of different task offloading schemes versus the number of IoUT devices.} \label{offload}
\end{figure}

To characterize the impact of the offloading scheme on the system's profit, in Fig. \ref{offload}, we show the profit of different task offloading schemes. The full offloading scheme represents that all IoUT devices select to offload their tasks to the surface-station for processing, while the non-offloading scheme means that all IoUT devices address their tasks locally. Moreover, the random offloading scheme represents that each device randomly chooses whether to offload their computing tasks to the surface-station, while the partial offloading scheme means that we designate a proportion of tasks to offload to the surface-station and leave some tasks to be processed locally. Although the non-offloading scheme can satisfy the requirements of the devices, the cost that it has to pay is substantially higher than that of offloading the tasks to the MTUC framework for processing due to the energy dissipation of IoUT devices that are difficult to recharge. The IoUT devices are also harder to recharge than the AUVs that can be continuously recharged. Similarly, the surface-stations may be more readily recharged. Furthermore, when we choose to offload some tasks to the MTUC, the profit gleaned increases significantly. Explicitly, the proposed scheme consistently outperforms the other offloading schemes because it can search for an optimal offloading strategy to maximize the profit with limited resources.

\begin{figure}
\centering
\includegraphics[width=7cm]{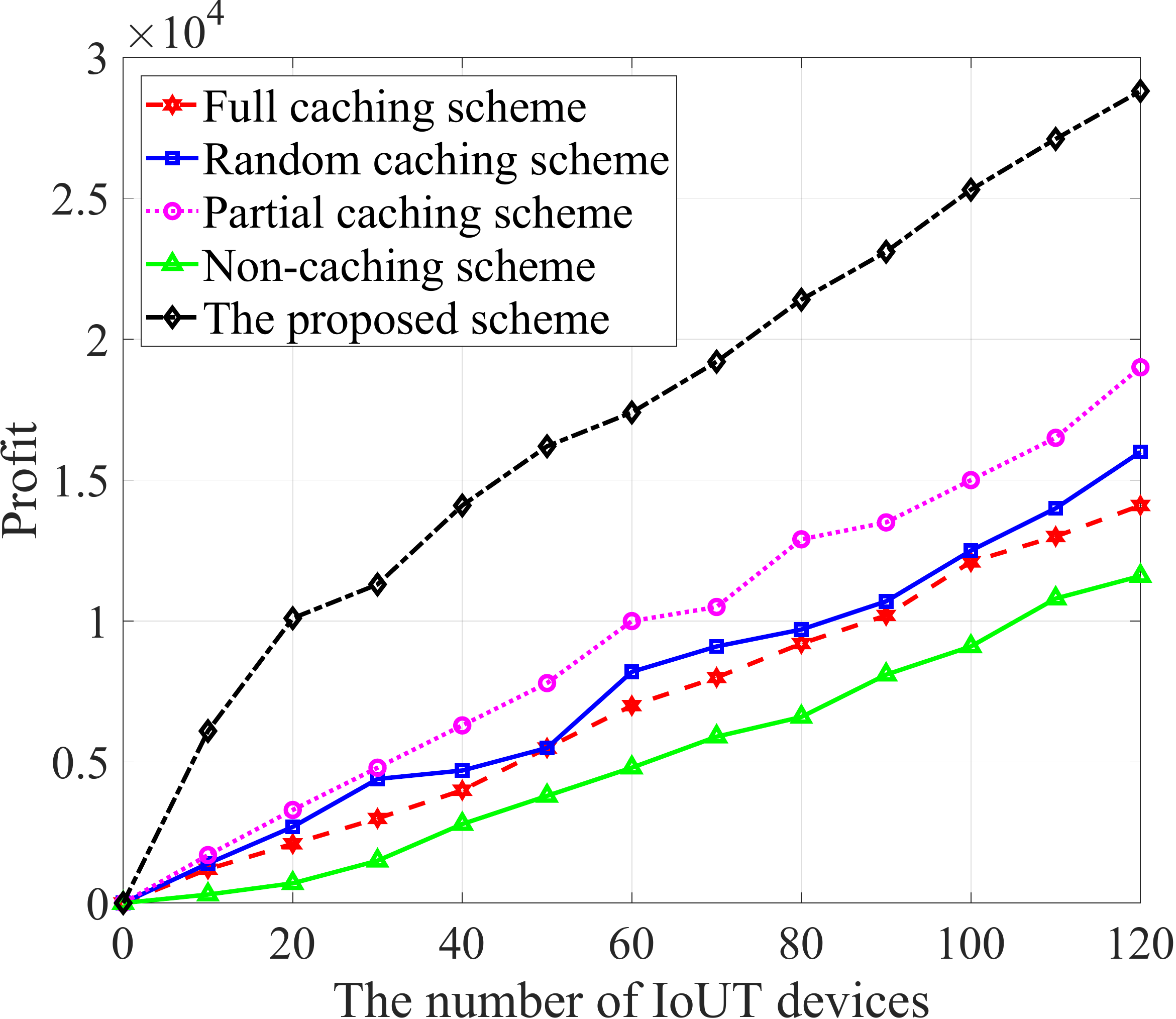}\\
\caption{The profit of different caching schemes versus the number of IoUT devices.} \label{cache}
\end{figure}

Fig. \ref{cache} shows the benefit of task caching. The full caching scheme represents that all tasks are cached on the  surface-station, while the non-caching scheme represents that none of the tasks is cached. Moreover, the random caching scheme means that we randomly choose some of the tasks to cache by the surface-station, while the partial caching scheme represents that we designate a certain proportion of tasks to cache by the surface-station. Firstly, we can observe that task caching significantly improves the system's profit, when there are repeated task computing requests. This is because task caching avoids repeated communication and computation, consequently reducing the processing latency and the energy consumption.  Furthermore, upon increasing the number of IoUT devices, the profit increases dramatically. The reason for this is that the more IoUT devices we have, the higher the probability of repeated computing requests. Moreover, we can see that the proposed scheme outperforms other schemes without optimization. This is because the scheme advocated comprehensively considers both the popularity of the tasks and the storage capacity of the surface-station for formulating an optimal caching strategy so as to attain the highest system profit.

\begin{figure}[t]
\centering
\includegraphics[width=7cm]{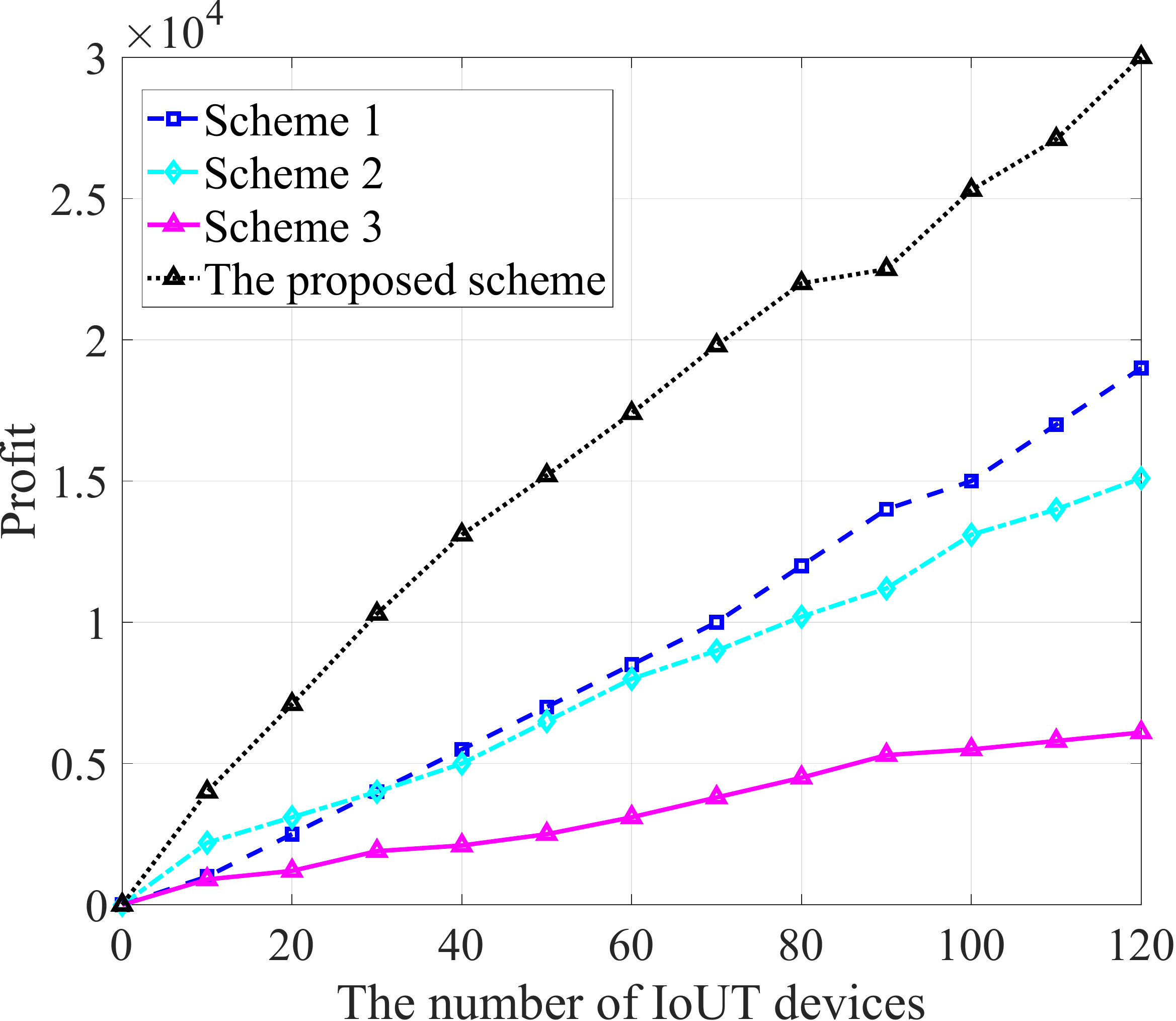}\\
\caption{The profit of different computing and communication allocation schemes versus the number of IoUT devices (Scheme 1 is with optimal bandwidth resource allocation and average computing resource allocation, scheme 2 is with average bandwidth resource allocation and optimal computing resource allocation, and scheme 3 is with average bandwidth resource allocation, and average computing resource allocation.).} \label{Band_Compute}
\end{figure}
To investigate the impact of computing and communication resource configuration on the system's profit, we compare the profit of different schemes in Fig. \ref{Band_Compute}. Observe that the scheme relying on the average bandwidth resource allocation and average computing resource allocation is the worst, because it ignores the differences in tasks and the resource states between different IoUT devices. By contrast, optimizing both the bandwidth resource allocation and computing resource allocation dramatically increases the system's profit. Furthermore, we can observe that the proposed scheme is much better than all other schemes that optimize a single resource individually, which indicates that the configuration of both types of resources significantly improves the system's profit.

\subsection{The Performance Analysis of The A3C Algorithm}


\begin{figure*}[t]\centering
\subfigure[3 AUV supporting 100 IoUT devices.]{\includegraphics[width=5.5cm]{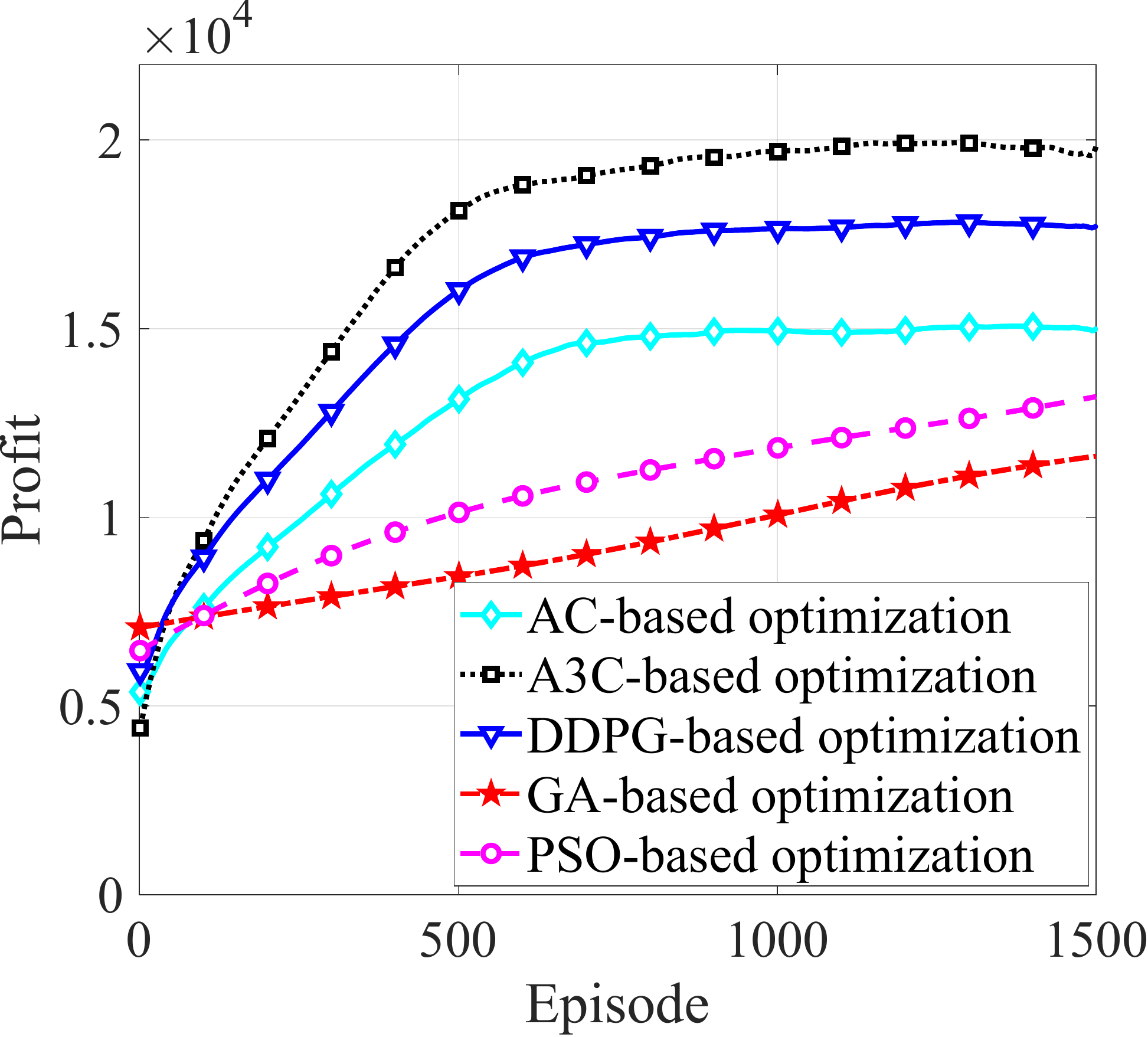}\label{Algorithm_comparison_M_1}}
\subfigure[4 AUV supporting 200 IoUT devices.]{  \includegraphics[width=5.5cm]{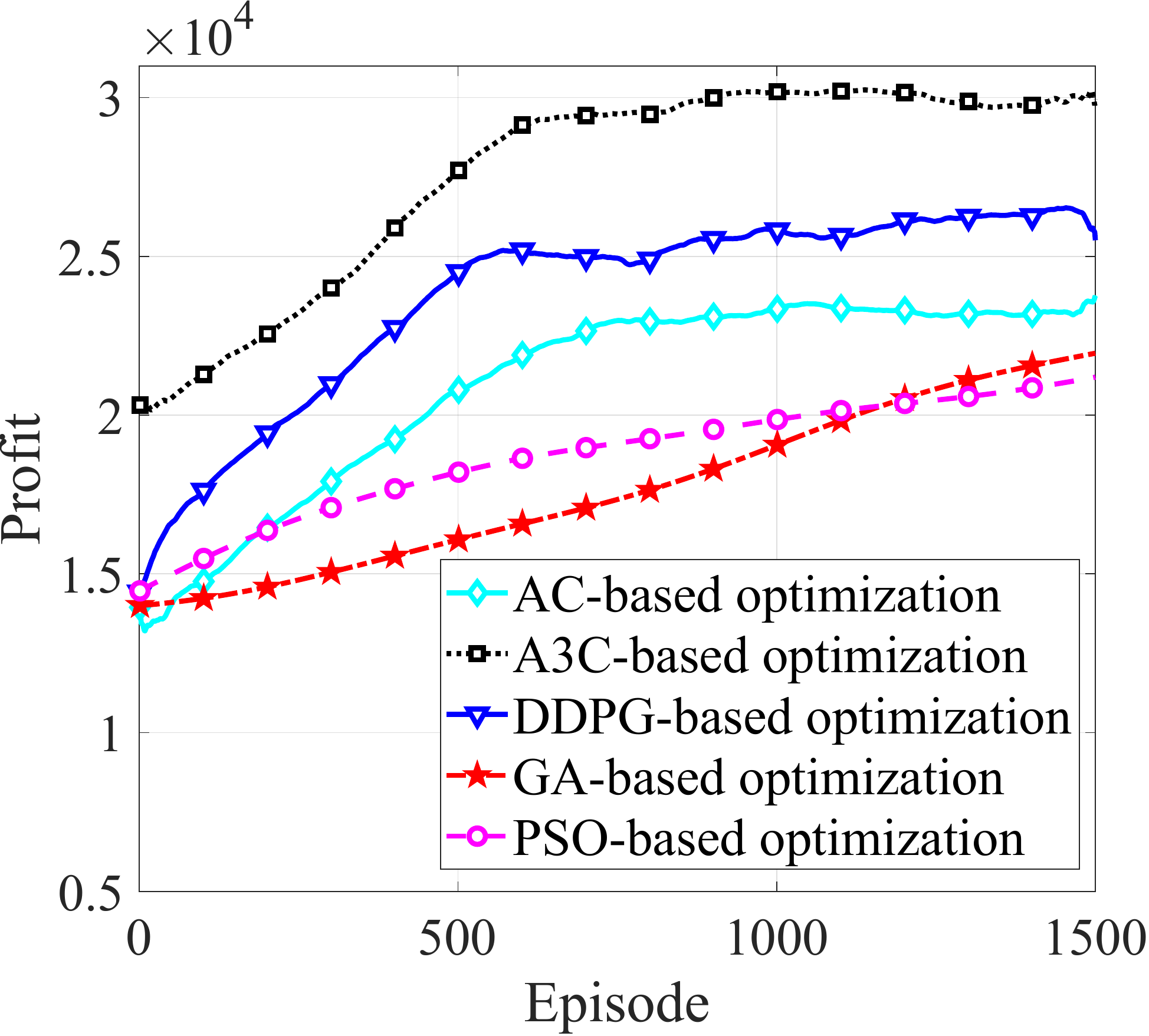} \label{Algorithm_comparison_M_2} }
\subfigure[5 AUV supporting 300 IoUT devices.]{  \includegraphics[width=5.5cm]{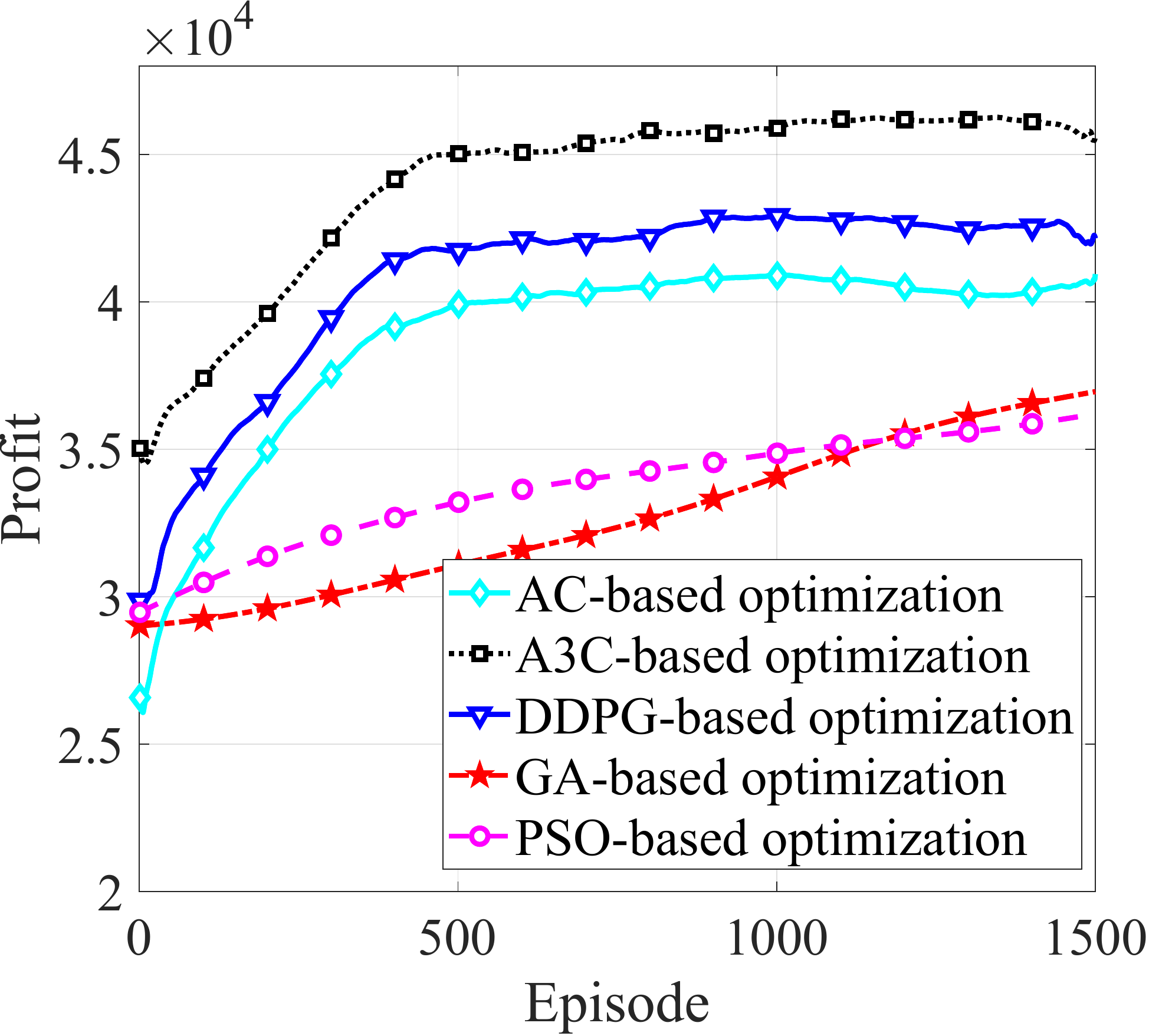} \label{Algorithm_comparison_M_3}}
\subfigure[6 AUV supporting 400 IoUT devices.]{  \includegraphics[width=5.5cm]{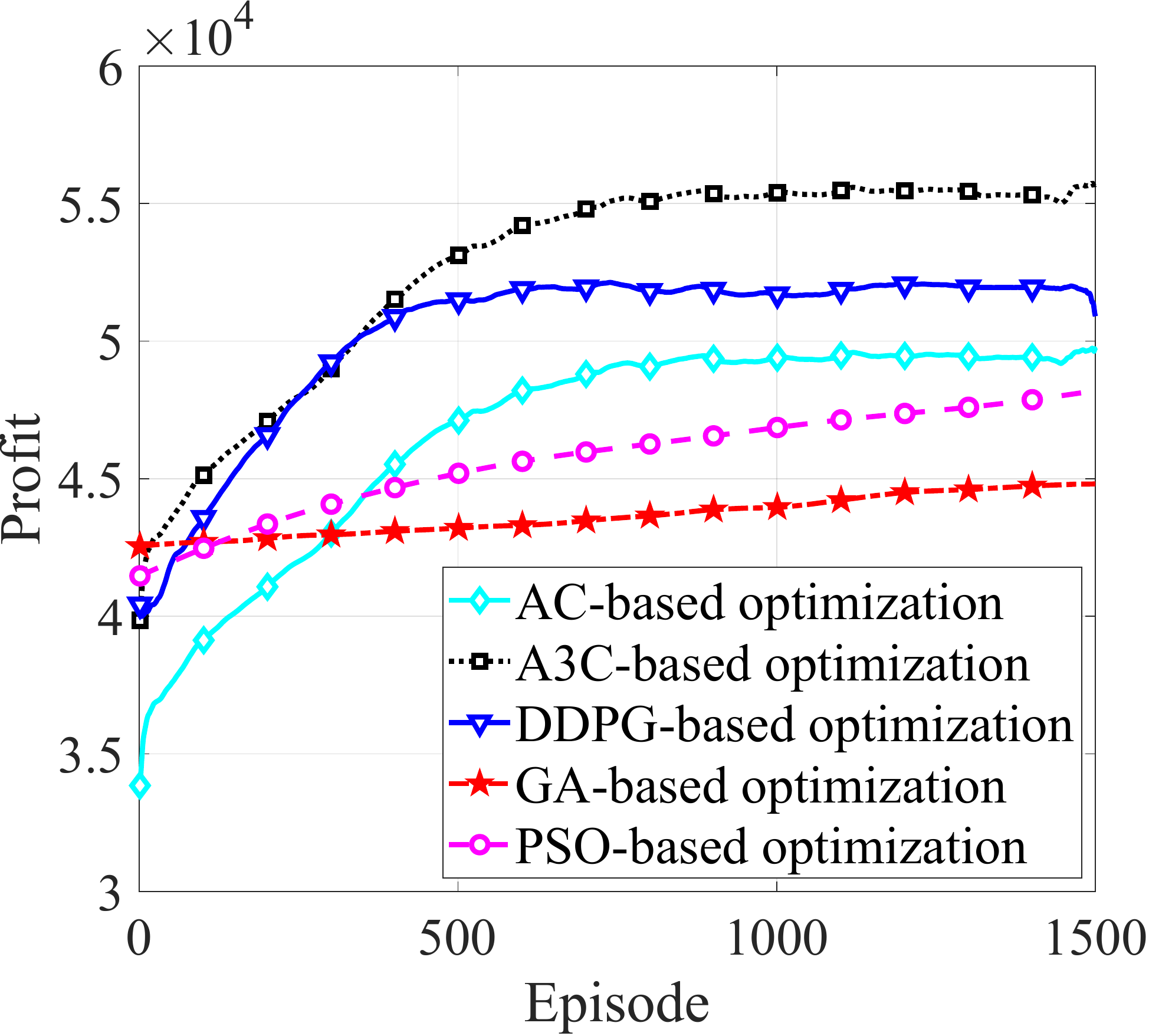}\label{Algorithm_comparison_M_4}  }
\subfigure[7 AUV supporting 500 IoUT devices.]{  \includegraphics[width=5.5cm]{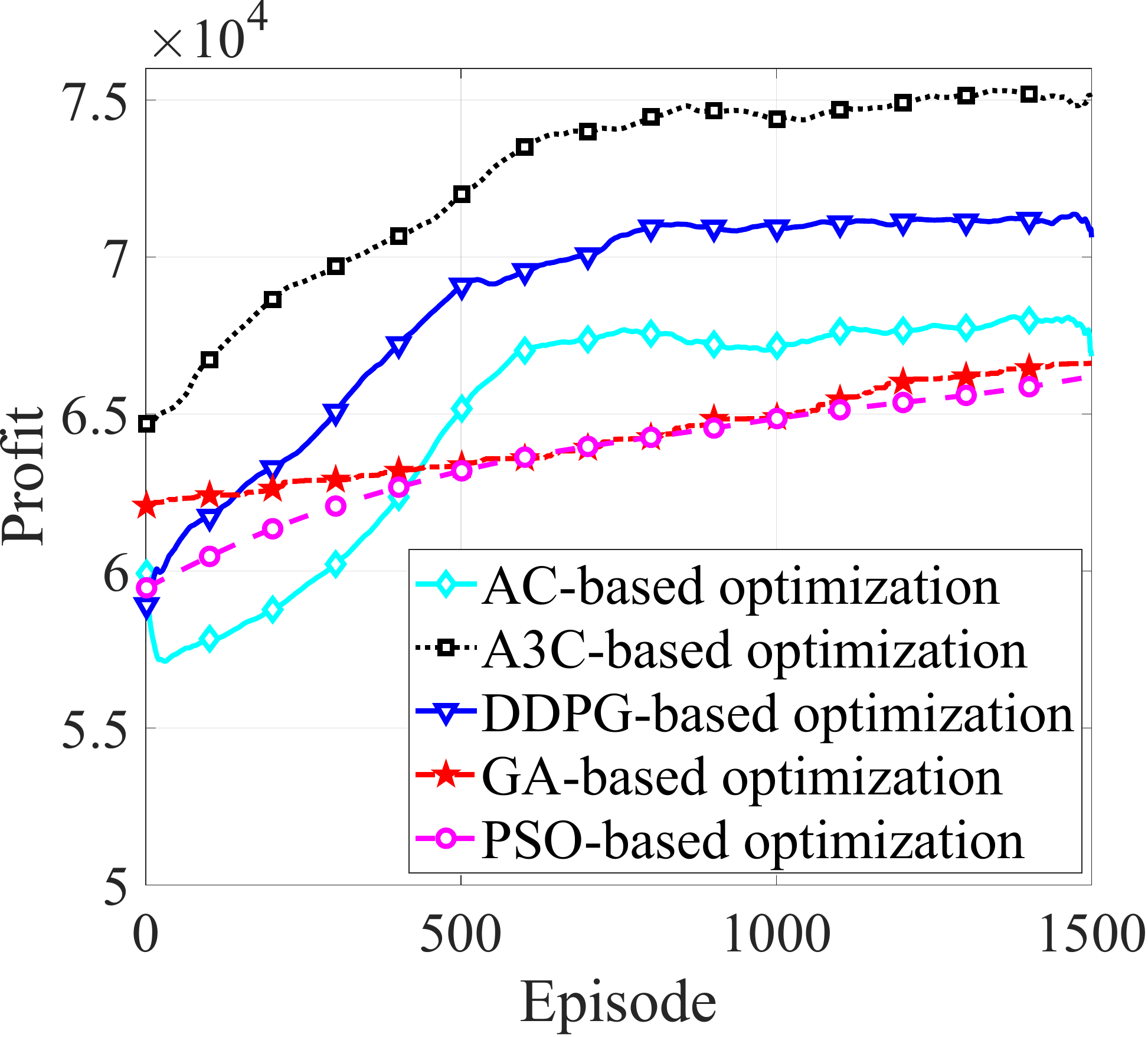} \label{Algorithm_comparison_M_5} }
\subfigure[8 AUV supporting 600 IoUT devices.]{  \includegraphics[width=5.5cm]{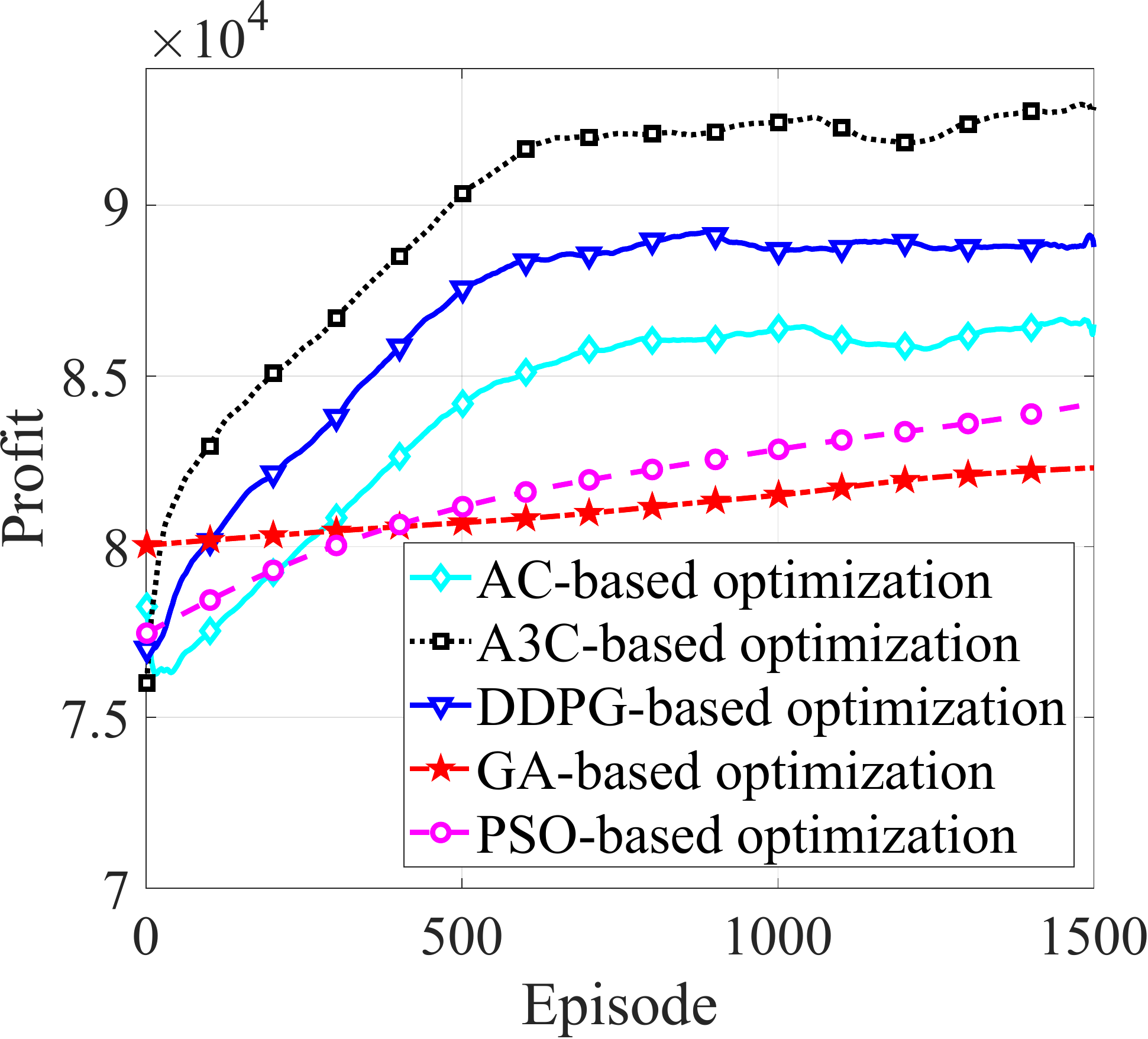} \label{Algorithm_comparison_M_6} }
 \caption{The profit of different algorithms.}\label{Algorithm_comparison_M}
\end{figure*}

Conventional methods falter in tackling $\mathcal{P}1$, because it is typically NP-hard and has a high dimensionality. In Fig. \ref{Algorithm_comparison_M}, we compare the performance of state-of-the-art algorithms in tackling this problem in multiple scenarios, including the popular genetic algorithm (GA) \cite{9075201}, particle swarm optimization (PSO) algorithm \cite{9373610}, actor-critic (AC) algorithm \cite{9363345}, deep deterministic policy gradient (DDPG) algorithm \cite{9427224}, and our A3C algorithm. Observe that the heuristic algorithms, i.e., GA and PSO-based optimization strategies have poor convergence performance. By contrast, the deep reinforcement learning algorithms, i.e., AC, DDPG, and A3C-based optimization strategies, perform better. The reason is that the deep reinforcement learning algorithms are more suitable for solving high-dimensional problems as a direct of the neural networks' powerful function fitting capability. Furthermore, the A3C algorithm is better than DDPG and AC-based optimization, because it can find better solutions within the same number of iterations as a benefit of its distributed parallel operating paradigm.

\begin{figure}[t]
\centering
\includegraphics[width=7cm,height=6.4cm]{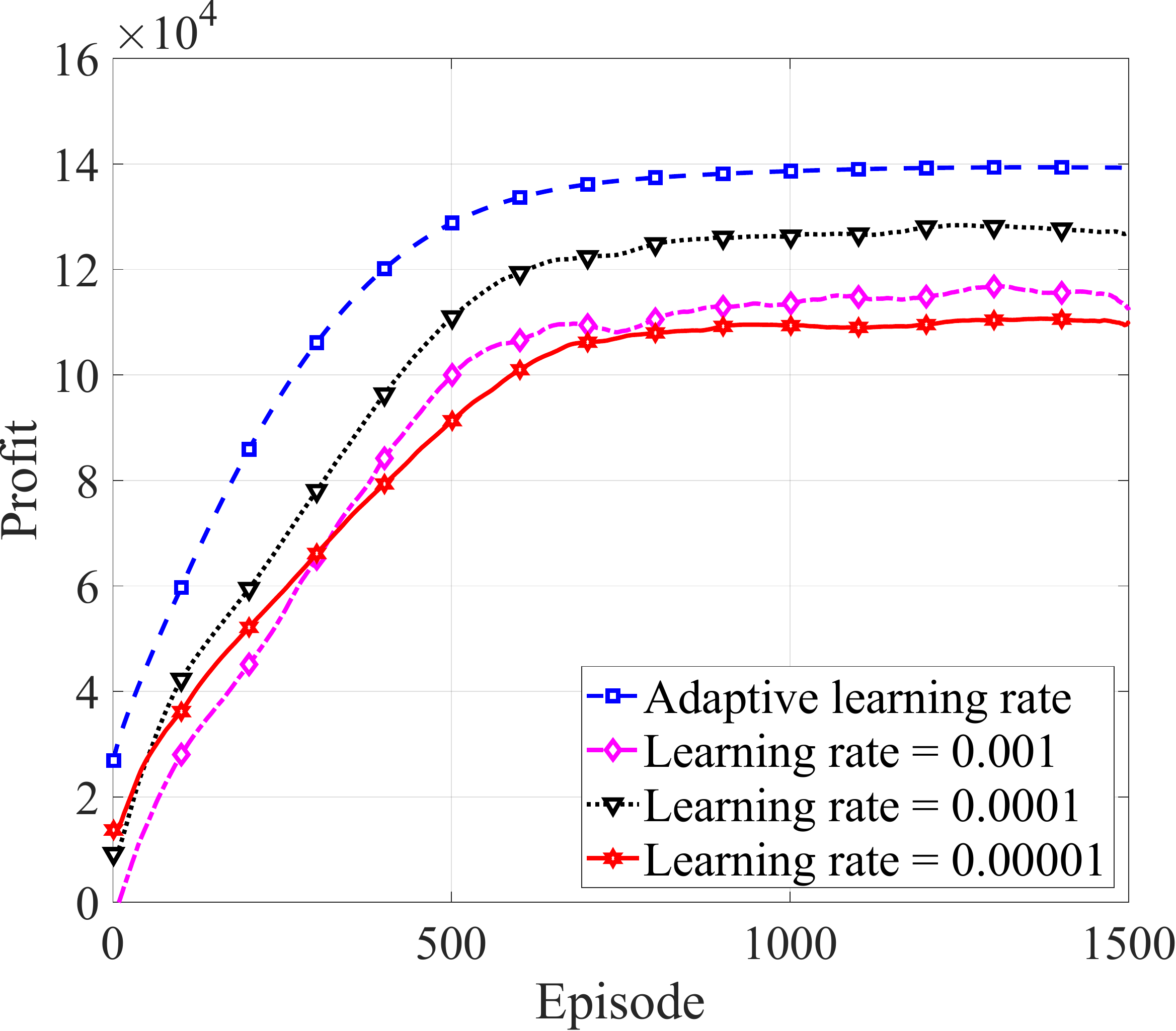}\\
\caption{Impact of the learning rate on the convergence performance of A3C.} \label{Learning_Rate}
\end{figure}
The setting of the hyperparameters in deep reinforcement learning is of pivotal importance, since it may seriously affect the performance of the algorithms. As a significant hyperparameter in A3C, the learning rate dramatically affects the convergence rate, but fails to obtain a theoretical optimal value. If the learning rate is set too low, it will slow down the convergence of the algorithm and increase the training time. By the contrast, if the learning rate is excessive, the parameters may swing back and forth on both sides of the optimal value, failing to converge. In Fig. \ref{Learning_Rate}, we investigate the impact of the learning rate on the convergence performance of A3C. As we can observe, the algorithm having an adaptive learning rate is superior to others, which will gradually adjust the learning rate according to the training process.

\section{Conclusions}
To satisfy the stringent requirements of  IoUT applications, we proposed an MTUC framework by judiciously allocating the computing, communication, and storage resources of both the surface-station, as well as of the AUVs, and of the IoUT devices. Furthermore, under this framework, we conceived a system-level optimization problem for the sake of maximizing the profit of the MTUC framework relying on jointly optimizing the environment-aware trajectory design of the AUVs, computation offloading, data caching, communication, and computing resource allocation. Since the problem formulated is NP-hard and of high dimensionality, we transformed it into an MDP and further employed the A3C algorithm to solve it. Finally, we conducted a range of experiments to validate the efficiency of the proposed scheme.

In the near future, we plan to study the impact of underwater environments on IoUT applications. For instance, hostile underwater environments may cause a high probability of device failure, reducing the success probability of IoUT applications. Hence it is beneficial to explore how to guarantee the success probability of IoUT applications. Moreover, the conception of having low communication overhead in the face of  limited UAC communication resources is also worth pursuing, for example by using federated learning.
\begin{appendices}
\setcounter{equation}{0}
\renewcommand\theequation{A.\arabic{equation}}
\section{Proof of Proposition 1}\label{ProofProposition1}
Let us consider a particular case, where the values of computation offloading strategy $\boldsymbol{O}$, caching strategy $\boldsymbol{H}$, bandwidth allocation $\boldsymbol{R}$ and computing resource allocation $\boldsymbol{F}$ are given, which satisfy the constraints demonstrated in problem $\mathcal{P}1$. Consequently, we can obtain a sub-problem of $\mathcal{P}1$ as
\begin{subequations}\label{OptimizationProblem}
\begin{eqnarray}\small
\label{63_sub}
\begin{aligned}
\mathcal{P}2: \quad & \underset{\boldsymbol{Y}}{\mathop{\max}} \quad \quad
 C-\!\sum\limits_{j=1}^{M}\chi{\hat{E}_j}\\
\end{aligned}
\end{eqnarray}
\begin{equation}\label{p1c10_sub}
s.t. \quad \sum\limits_{j=1}^{M}{\sum\limits_{\xi=1}^{{{S}_{j}}}{{{Y}_{{{j}_{k}}}}[\xi]}}=1, \forall j \in \boldsymbol{M},  \forall k \in \boldsymbol{K},
\end{equation}
\begin{equation}\label{p1c9_sub}
\boldsymbol{P}_{j}^{\textrm{A}}[S_j+1]=\boldsymbol{P}_{j}^{\textrm{A}}[0], \forall j \in \boldsymbol{M},
\end{equation}
\begin{equation}\label{p1c11_sub}
\sum\limits_{\xi=1}^{{{S}_{j}}}{\sum\limits_{k=1}^{K}{{{Y}_{{{j}_{k}}}}}}[\xi]={{S}_{j}}, \forall j \in \boldsymbol{M},
\end{equation}
\begin{equation}\label{p1c12_sub}
\sum\limits_{j=1}^{M}{{{S}_{j}}}=K, \forall j \in \boldsymbol{M},
\end{equation}
\begin{equation}\label{p1c13_sub}
{{{T}}^{\textrm{AT}}_{\max }}-{{T}^{\textrm{AT}}_{\min }}\le \varepsilon.
\end{equation}
\end{subequations}
where $C$ is a constant associated with the first term of Eq. (\ref{63}), while ${\hat{E}_j}$ is represented as
\begin{eqnarray}
\label{38_proof}
\hat{E}_{j}=\sum\limits_{k=1}^{K}{\sum\limits_{\xi=1}^{{{S}_{j}}}{{{Y}_{{{j}_{k}}}}[\xi]{{\hat{A}}_{k}}}P_j^{\rm H}[\xi] }+{\sum\limits_{\xi=0}^{{{S}_{j}}}\frac{{{{d}_{j}}[\xi]}}{V_{k}}P_j^{\rm F}[\xi]  },
\end{eqnarray}
where ${{\hat{A}}_{k}}$ is a constant as a result of Eq.  (\ref{29}) after the computation offloading strategy $\boldsymbol{O}$, caching strategy $\boldsymbol{H}$, bandwidth allocation $\boldsymbol{R}$ and computing resource allocation $\boldsymbol{F}$ are given. In fact, problem $\mathcal{P}2$ can be equivalent to
\begin{subequations}\label{OptimizationProblem}
\begin{eqnarray}\small
\label{63_sub_eq}
\begin{aligned}
\mathcal{P}3: \quad & \underset{\boldsymbol{Y}}{\mathop{\min}}
\!\sum\limits_{j=1}^{M}\sum\limits_{k=1}^{K}{\sum\limits_{\xi=1}^{{{S}_{j}}}{{{Y}_{{{j}_{k}}}}[\xi]{{\hat{A}}_{k}}}P_j^{\rm H}[\xi] }+{\sum\limits_{\xi=0}^{{{S}_{j}}}\frac{{{{d}_{j}}[\xi]}}{V_{k}}P_j^{\rm F}[\xi]  }\\
\end{aligned}
\end{eqnarray}
\begin{equation}
s.t. ~~ \textrm{Eq. (\ref{p1c10_sub}) $\sim$  (\ref{p1c13_sub})}.
\end{equation}
\end{subequations}

$\mathcal{P}3$ can be seen as a variant of the multiple traveling salesman problem (MTSP), which is essentially a generalization of the well-known traveling salesman problem (TSP). Furthermore, since TSP has already been proven to be NP-hard and can be reduced to the MTSP, MTSP is an NP-hard problem  \cite{cheikhrouhou2021comprehensive}. Consequently, $\mathcal{P}3$ is NP-hard. Furthermore, due to $\mathcal{P}3$ is a sub-problem of $\mathcal{P}1$, we can determine that $\mathcal{P}1$ is also an NP-hard problem. Therefore, if $P \neq NP$, there is no algorithm can solve $\mathcal{P}1$ in polynomial time. Thus the proof  of \emph{Proposition \ref{proposition1}} is completed.
\end{appendices}

\bibliographystyle{IEEEtran}
\bibliography{ref}
\vspace{-5mm}
\begin{IEEEbiography}[{\includegraphics[width=1in,height=1.33in]{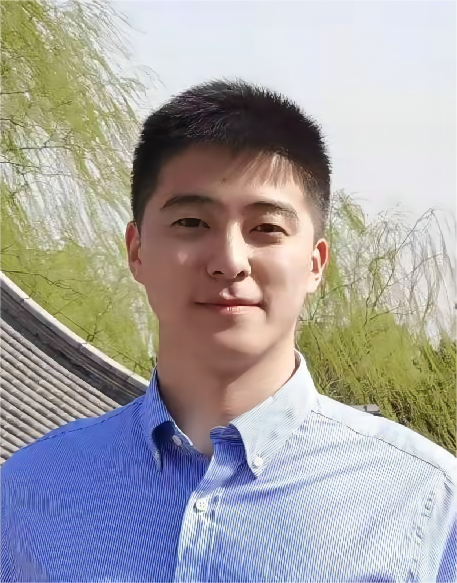}}]{\textbf{Xiangwang Hou}} (Student Member, IEEE) is currently pursuing his Ph.D. degree in Electronics and Communication Engineering at Tsinghua University, Beijing, China. And he received the B.E. degree in Electronic Information Engineering from Shandong University of Technology, Shandong, China in 2017 and the M.E. degree in Information and Communication Engineering from Xidian University, Xi'an, China in 2020.  His research interests include UAV/AUV networks, federated learning and wireless AI.
 \vspace{-6mm}
  \end{IEEEbiography}
  \begin{IEEEbiography}[{\includegraphics[width=1in,height=1.33in]{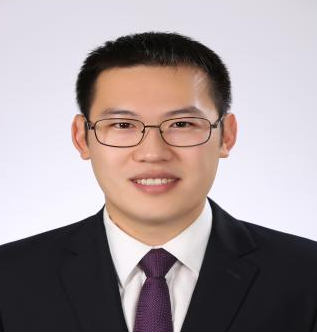}}]{\textbf{Jingjing Wang}} (Senior Member, IEEE) received his B.S. degree in Electronic Information Engineering from Dalian University of Technology, Liaoning, China in 2014 and the Ph.D. degree in Information and Communication Engineering from Tsinghua University, Beijing, China in 2019, both with the highest honors. From 2017 to 2018, he visited the Next Generation Wireless Group chaired by Prof. Lajos Hanzo, University of Southampton, UK. Dr. Wang is currently an associate professor at School of Cyber Science and Technology, Beihang University. His research interests include AI enhanced next-generation wireless networks, UAV swarm intelligence and confrontation. He has published over 100 IEEE Journal/Conference papers. Dr. Wang was a recipient of the Best Journal Paper Award of IEEE ComSoc Technical Committee on Green Communications \& Computing in 2018, the Best Paper Award of IEEE ICC and IWCMC in 2019.
    \end{IEEEbiography}
\begin{IEEEbiography}[{\includegraphics[width=1in,height=1.33in]{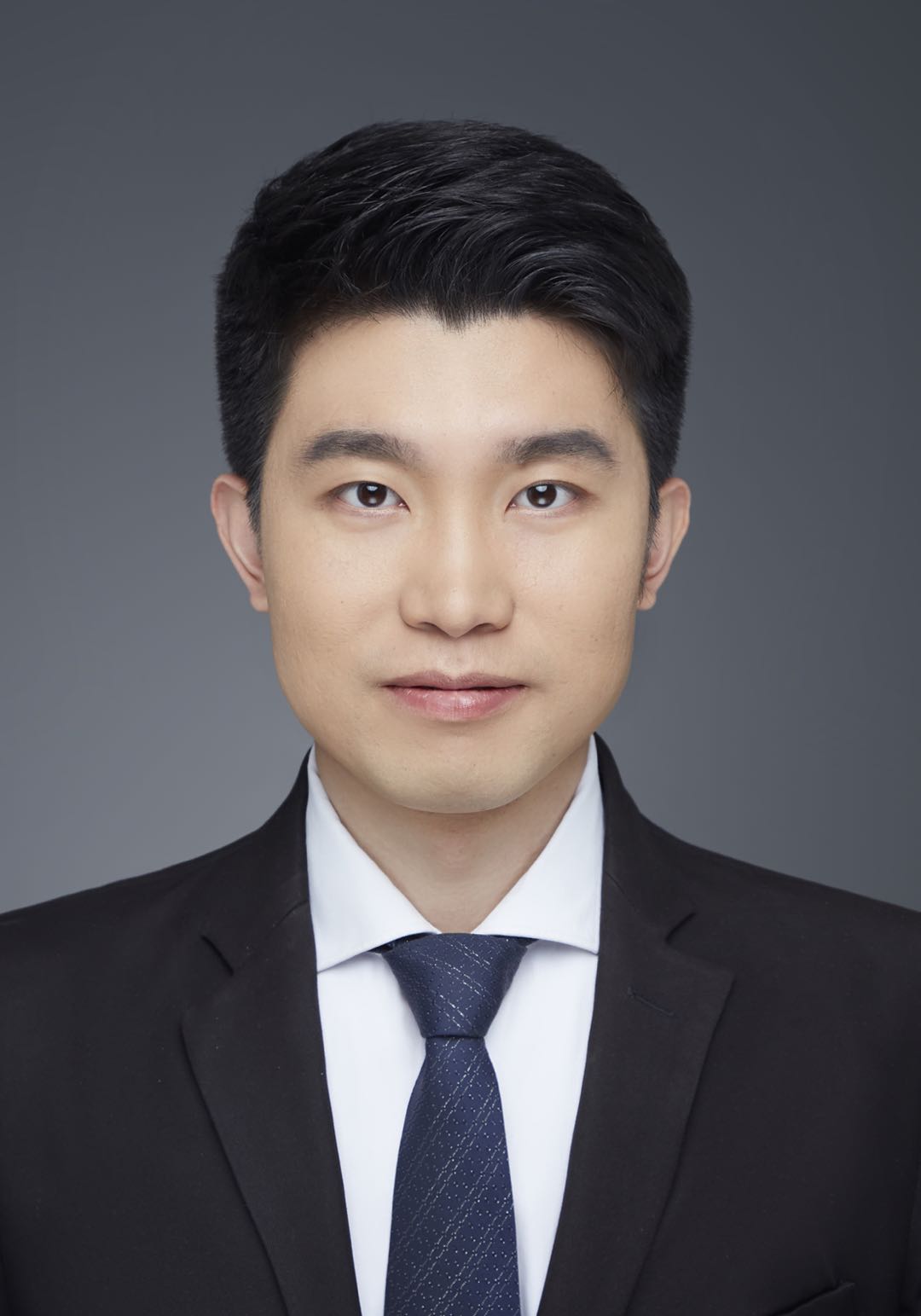}}]{\textbf{Tong Bai}} (Member, IEEE) received the B.Sc. degree in telecommunications from Northwestern Polytechnical University, Xi'an, China, in 2013, and the M.Sc. and Ph.D. degrees in communications and signal processing from the University of Southampton, Southampton, U.K., in 2014 and 2019, respectively. From 2019 to 2020, he was a Postdoctoral Researcher with Queen Mary University of London, London, U.K.  Since 2020, he has been with Beihang University (BUAA) as an Assistant Professor. His research interests include edge intelligence and wireless communications.
\end{IEEEbiography}
\begin{IEEEbiography}[{\includegraphics[width=1in,height=1.33in]{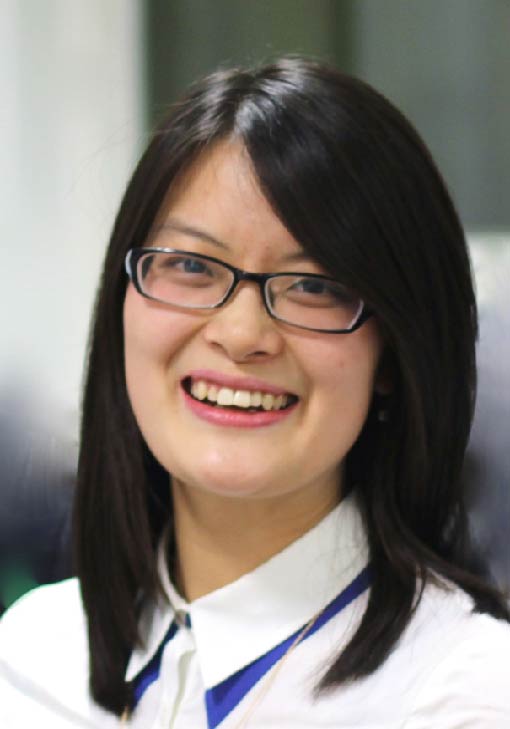}}]{\textbf{Yansha Deng}} (Senior Member, IEEE) received the Ph.D. degree in electrical engineering from the Queen Mary University of London, U.K., in 2015. From 2015 to 2017, she was a Post-Doctoral Research Fellow with King’s College London, U.K, where she is currently a Senior Lecturer (an Associate Professor) with the Department of Engineering. Her research interests include molecular communication and machine learning for 5G/6G wireless networks. She was a recipient of the Best Paper Awards from ICC 2016 and GLOBECOM 2017 as the first author and IEEE Communications Society Best Young Researcher Award for the Europe, Middle East, and Africa Region 2021. She also received the Exemplary Reviewers of the IEEE Transactions on communications in 2016 and 2017 and IEEE Transactions on wireless communications in 2018. She has served as a TPC Member for many IEEE conferences, such as IEEE GLOBECOM and ICC. She is currently an Associate Editor of the IEEE Transactions on communications and IEEE Transactions on molecular, biological and multi-scale communications, a Senior Editor of the IEEE communication letters, and the Vertical Area Editor of IEEE Internet of things magazine.
\end{IEEEbiography}
 \begin{IEEEbiography}[{\includegraphics[width=1in,height=1.33in]{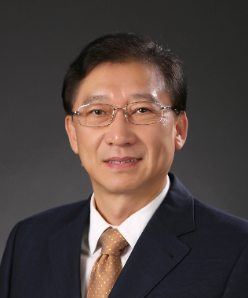}}]{\textbf{Yong Ren}} (Senior Member, IEEE) received his B.S, M.S and Ph.D. degrees in electronic engineering from Harbin Institute of Technology, China, in 1984, 1987, and 1994, respectively. He worked as a post doctor at Department of Electrical Engineering, Tsinghua University, China from 1995 to 1997. Now he is a full professor of Department of Electronic Engineering and serves as the director of the Complexity Engineered Systems Lab in Tsinghua University. Moreover, he is also a guest professor of the Network and Communication Research Center in Peng Cheng Laboratory. He has authored or co-authored more than 400 technical papers in the area of computer network and mobile telecommunication networks. He has served as a reviewer of more than 40 international journals or conferences. His current research interests include marine information network, swarm intelligence and wireless AI.
 \end{IEEEbiography}
  \begin{IEEEbiography}[{\includegraphics[width=1in,height=1.33in]{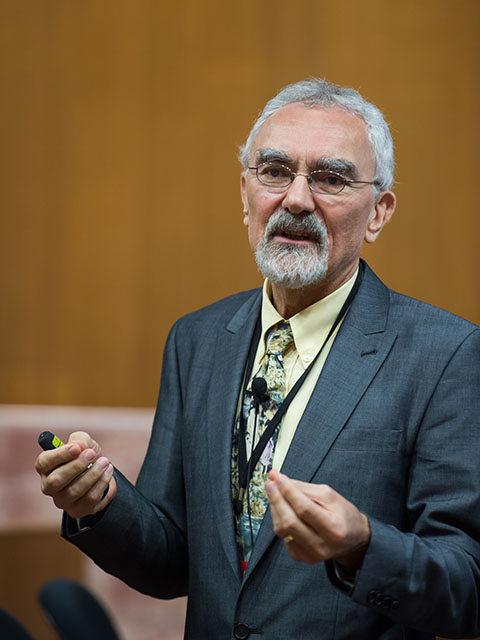}}]{\textbf{Lajos Hanzo}} (Life Fellow, IEEE)  (\url{http://www-mobile.ecs.soton.ac.uk}, \url{https://en.wikipedia.org/wiki/Lajos_Hanzo}) received his Master degree and Doctorate in 1976 and 1983, respectively from the Technical University (TU) of Budapest. He was also awarded the Doctor of Sciences (DSc) degree by the University of Southampton (2004) and Honorary Doctorates by the TU of Budapest (2009) and by the University of Edinburgh (2015).  He is a Foreign Member of the Hungarian Academy of  Sciences and a former Editor-in-Chief of the IEEE Press.  He has served several terms as Governor of both IEEE ComSoc and of VTS.  He has published 2000+ contributions at IEEE Xplore, 19 Wiley-IEEE Press books and has helped the fast-track career of 123 PhD students. Over 40 of  them are Professors at various stages of their careers in academia and many of them are leading scientists in the wireless industry. He is also a Fellow of the Royal Academy of Engineering (FREng), of the IET and of EURASIP. He is the recipient of the 2022 Eric Sumner Field Award.
  \end{IEEEbiography}
\end{document}